\documentclass[12pt]{article}
\usepackage[pdftex]{graphicx}
\usepackage[normalem]{ulem}

\usepackage{amsmath}
\usepackage[svgnames,hyperref]{xcolor}
\usepackage{empheq}
\usepackage[svgnames,hyperref]{xcolor}
\usepackage[skins,theorems,many]{tcolorbox}
\usepackage{bm}
\usepackage{hyperref}
\usepackage{caption}
\usepackage[font=footnotesize,labelfont=bf]{caption}
\usepackage{algorithm}
\usepackage[noend]{algpseudocode}
\usepackage{calrsfs}
\usepackage[svgnames,hyperref]{xcolor}
\usepackage{empheq}
\usepackage[svgnames,hyperref]{xcolor}
\usepackage[skins,theorems,many]{tcolorbox}
\usepackage{bm}
\usepackage{hyperref}
\usepackage{caption}
\usepackage{algorithm}
\usepackage[noend]{algpseudocode}
\usepackage{calrsfs}
\usepackage{amssymb}
\usepackage{color,wrapfig}
\usepackage{soul}
\usepackage{dirtytalk}
\usepackage{cancel}

\usepackage{soul}
\usepackage{color}

\usepackage{subcaption}
\newcommand{\figuremacro}[5]{
    \begin{figure}[#1]
        \centering
        \includegraphics[width=#5\columnwidth]{#2}
        \caption[#3]{\textbf{#3}#4}
        \label{fig:#2}
    \end{figure}
}

\newcommand{\beginsupplement}{%
        \setcounter{table}{0}
        \renewcommand{\thetable}{S\arabic{table}}%
        \setcounter{figure}{0}
        \renewcommand{\thefigure}{S\arabic{figure}}%
     }
     
\begin{document}

\pagenumbering{arabic}

\begin{center}
{\large \bf{The mechanical response of fire ant rafts}}
\end{center}

\begin{center}
 {\bf Robert J. Wagner$^1$, Samuel Lamont$^2$, Zachary T. White$^2$, and Franck J. Vernerey$^2$
 \footnote{correspondence to Franck.Verenerey@Colorado.edu}}\\ 
\end{center}

\begin{center}
$^1$Sibley School of Mechanical \& Aerosspace Engineering, \\
Cornell University, Ithaca, NY, USA\\
$^2$Paul M. Rady School of Mechanical Engineering, \\
University of Colorado, Boulder, CO, USA \\
\end{center}

\textit{Fire ants (Solenopsis invicta) cohesively aggregate via the formation of voluntary ant-to-ant attachments when under confinement or exposed to water. Once formed, these aggregations act as viscoelastic solids due to dynamic bond exchange between neighboring ants as demonstrated by rate-dependent mechanical response of 3D aggregations, confined in rheometers. We here investigate the mechanical response of 2D, planar ant rafts roughly as they form in nature. Specifically, we load rafts under uniaxial tension to failure, as well as to 50$\%$ strain for two cycles with various recovery times between. We do so while measuring raft reaction force (to estimate network-scale stress), as well as the networks' instantaneous velocity fields and topological damage responses to elucidate the ant-scale origins of global mechanics. The rafts display brittle-like behavior even at slow strain rates (relative to the unloaded bond detachment rate) for which Transient Network Theory predicts steady-state creep. This provides evidence that loaded ant-to-ant bonds undergo mechanosensitive bond stabilization or act as \say{catch bonds}. This is further supported by the coalescence of voids that nucleate due to biaxial stress conditions and merge due to bond dissociation. The characteristic timescales of void coalescence due to chain dissociation provide evidence that the local detachment of stretched bonds is predominantly strain- (as opposed to bond lifetime-) dependent, even at slow strain rates, implying that bond detachment rates diminish significantly under stretch. Significantly, when the voids are closed by restoring the rafts to unstressed conditions, mechanical recovery occurs, confirming the presence of concentration-dependent bond association that - combined with force-diminished dissociation - could further bolster network cohesion under certain stress states.}

\section{Introduction}

\say{Dynamic bonds} that break and reform without damage are prerequisite to the spontaneous functionality and complex mechanical response of living systems across length scales spanning from protein folding \cite{gallivan_cation-_1999}, mitosis \cite{rincon_kinesin-5-independent_2017}, and tissue healing \cite{ajeti_wound_2019}, to the reconfiguration of superorganismal networks \cite{vicsek_collective_2012,hu_entangled_2016}. Elucidating the mechanosensitive kinetics of such bonds is an essential step toward understanding, and therefore mimicking, the emergent mechanical traits they yield in collective structures. One class of living materials whose properties are intrinsically governed by such kinetics, are entanglements of insects \cite{peleg_collective_2018,tennenbaum_mechanics_2016} and worms \cite{deblais_rheology_2020}. Amongst such systems are aggregations comprised entirely of fire ants (\textit{S. invicta}) \cite{wagner_treadmilling_2021,wagner_computational_2022}, which coalesce into roughly planar raft structures during floods \cite{mlot_fire_2011}, via the formation of physical ant-to-ant (e.g., leg-to-leg, leg-to-mandible, etc.) connections \cite{foster_fire_2014}, which are herein referred to as `bonds'. This evolutionary propensity to grab onto one another for flood survival can be exploited to manually create bulk ant aggregations, either by placing them in water or perturbed confinement. Such aggregations display intricate morphological and mechanical behaviors ranging from tower- \cite{phonekeo_fire_2017} and floating bridge-building \cite{wagner_treadmilling_2021,wagner_computational_2022} to non-Newtonian \cite{tennenbaum_mechanics_2016,vernerey_how_2018} and active rheological \cite{tennenbaum_activity_2020} responses. While the mechanical behavior of artificially-induced 3D ant aggregations has become fairly well studied \cite{tennenbaum_mechanics_2016,vernerey_how_2018,phonekeo_ant_2016}, surprisingly little remains known about the dynamic bonding and consequential rate-dependent mechanics of roughly planar \cite{mlot_fire_2011,wagner_treadmilling_2021} ant rafts approximately as they exist in nature, despite their rich morphological response in both unperturbed conditions \cite{wagner_treadmilling_2021,wagner_computational_2022,tennenbaum_activity_2020,anderson_janssen_2022,anderson_ant_2023,phonekeo_fire_2017} and within water currents \cite{ko_fire_2022}. 

In simple flooded conditions, fire ant rafts are relatively thin structures composed of two interacting layers \cite{adams_raft_2011,mlot_fire_2011,wagner_treadmilling_2021,wagner_computational_2022}. The bottom layer, here denoted as the structural layer, is made of a connected network of agents (the ants) that can transmit physical forces through the raft via ant-to-ant bonds. The top layer is made of a dispersed group of freely active ants that walk on top of the bottom layer and contribute to the raft morphogenesis over minute-to-hour timescales \cite{wagner_treadmilling_2021}. While this top layer may play a role in ant rafts' long-term self-healing (or dispersion) capabilities, we here focus on the mechanics of the structural network comprising the bulk of the rafts. Recently, we discovered that in unperturbed conditions these structural networks perpetually contract, yet their overall raft areas are approximately conserved over minute-to-hour timescales due to the deposition of new network structure at their perimeters \cite{wagner_treadmilling_2021}. We further discovered that both of these processes are enabled by ants' state transitions from the condensed structural layer to the dispersed top layer of freely walking ants, and \textit{vice versa}. These mechanisms corroborate the existence of dynamic bond exchange in ant rafts at timescales on the order of 10$^2$-10$^3$ seconds. Since then, Ko \textit{et al.} (2022) \cite{ko_small_2022,ko_fire_2022} have confirmed that small rafts are weakly cohered by both ant-to-ant bonds and surface tension \cite{ko_small_2022}, and undergo morphological change due to fluid flow-induced shear \cite{ko_fire_2022}. 

In this work, we conduct mechanical testing on experimental ant rafts to further understand how network topology and the dynamic bond kinetics affect the emerging mechanics of the raft. To probe rafts' rate-dependent responses and self-healing, we loaded them under uniaxial tension to failure at variable strain rates, as well as over two loading cycles to 50\% strain with variable recovery times between, while measuring the reaction force and estimating raft-scale stress (see \textbf{Section \ref{sec: Methods: Experimental Setup}-\ref{sec: Methods: Strain Rate Selection}}). We find that the mechanical properties of a raft in its natural state are in sharp contrast to those measured for confined 3D aggregtes using a rheometer \cite{tennenbaum_mechanics_2016,vernerey_how_2018}. Additionally, we characterized the topological signature of the structural networks using custom image analysis and particle image velocimetry (PIV). Experimental results indicate that ant rafts behave more elastically under applied strain than they do in unperturbed conditions -- a behavior suggestive of mechanosensitive catch-bond kinetics akin to those found in some biological, molecular systems \cite{pullen_catch_2017,sokurenko_catch_2008,thomas_biophysics_2008,tabatabai_detailed_nodate}. Furthermore, nucleation and coalescence of voids occur in strained rafts due to the finite length of leg-to-leg connections. When returned to their initial shapes, ant rafts are able to self-heal void defects that formed during deformation and, therefore, recover mechanical strength, which is attributed to density-dependent association kinetics. Together, these effects culminate in a material that is effectively a transient network when at rest, but a damage-prone elastic network under load. Despite this, some degree of rate-dependent mechanical response is preserved due to rafts' rates of conformational change, analogous to $\alpha$-relaxation in polymers \cite{rubinstein_polymer_nodate}.

\section{Rate-dependent mechanical response due to conformational relaxation} 
\label{sec: Results: Conformational Response}

Dynamic bonds' ability to break from stressed configurations and reattach into lower energy states enables dissipative flow that typically toughens materials, enhances their extensibilities, and imbues them with intrinsic rate-dependent responses \cite{tanaka_viscoelastic_1992,vernerey_statistically-based_2017,wagner_network_2021}. In networks where the characteristic bond detachment rate, $k_d$, is significantly slower than the attachment rate, $k_a$, the relaxation time imparted by dynamic bond exchange is roughly $k_d^{-1}$ \cite{vernerey_statistically-based_2017}. For both biological and synthetic dynamic polymers comprised of molecular constituents (e.g., cytoskeletal actin networks \cite{lieleg_cytoskeletal_2009} or epoxy vitrimers \cite{hubbard_creep_2022}) when the applied strain rate, $\dot \epsilon$, greatly exceeds the nominal detachment rate, $k_d$, these networks tend to behave elastically and exhibit brittle-like fracture, while when $\dot \epsilon \ll k_d$, these materials tend to exhibit viscous creep. Therefore, the strain rate is often normalized by the characteristic bond detachment rate, defining the dimensionless Weissenberg number $W=\dot \epsilon/k_d$, which provides a better intuition of the conditions under which a dynamic network is loaded. In particular, the Transient Network Theory \cite{vernerey_statistically-based_2017,wagner_network_2021} predicts that under uniaxial loading conditions, the creep-to-fracture transition occurs for a critical Weissenberg number around 0.5. Therefore, provided clear evidence of dynamic bonding in fire ant structures \cite{tennenbaum_mechanics_2016,phonekeo_ant_2016,vernerey_how_2018,wagner_treadmilling_2021}, we hypothesized that bond exchange would allow ant rafts to undergo or exhibit close-to steady-state creep when loaded at slow strain rates (relative to their unperturbed or \say{nominal} bond dissociating rate, $\bar k_d^0$), while displaying increasingly elastic behavior at higher relative loading rates.

To test this hypothesis, we loaded ant rafts in uniaxial tension to failure at strain rates of $\dot \epsilon = \{0.3, 1, 2, 4, 6\}$\% Hz. For details of the experimental set-up and procedure, see \textbf{Sections \ref{sec: Methods: Experimental Setup}-\ref{sec: Methods: Force Estimation}}, and \textbf{Figs. \ref{SI: Exp setup}-\ref{SI: Force calibration}}. Based on the distribution of attached bond lifetimes, we estimate that the nominal dissociation rate is $\bar k_d^0 \approx 7.4$ mHz so that these strain rates correspond to Weissenberg numbers of $W=\{0.45,1.35,2.70,5.41,8.11\}$, respectively.\footnote{Note that the value of $\bar k_d^0$ measured here is two orders of magnitude lower than the value, $k_d \approx 1.4$ Hz, predicted by the statistical mechanics model of Vernerey, \textit{et al.} (2018), which was developed for 3D aggregations undergoing parallel plate shear \cite{vernerey_how_2018}. This is likely due to the fact that the 3D aggregations' effective \say{crosslinks} or bonding modes are inclusive of leg entanglements and friction, as evidenced by the fact that aggregations of dead ants - incapable of forming the types of intentional bonds formed in the 2D raft structures investigated here - exhibited comparable mechanical properties to living aggregations within a certain strain rate regime \cite{tennenbaum_mechanics_2016}. Such entanglements and friction points likely express a much higher nominal dissociation rate.} For details on nominal dissociation and association kinetics, see \textbf{Section \ref{sec: Methods: Estimating Unperturbed Bond Dynamic Rates}} and \textbf{Fig. \ref{SI: Bond dynamics}}. For details on strain rate selection, refer to \textbf{Section \ref{sec: Methods: Strain Rate Selection}} and \textbf{Figs. \ref{SI: Active contraction}-\ref{SI: Delamination}}. The ensemble average ($n=4$) engineering stress-strain response, presented in \textbf{Fig. \ref{fig: stress-strain}.A}, indicates that ant rafts are very soft structures with peak stresses on the order of just 100 Pa. Despite the low-stress magnitude, their mechanical response exhibits measurable rate-dependence, whereby higher strain rates produced higher peak stresses, $T_u$ (\textbf{Fig. \ref{fig: stress-strain}.D}), and initial tangent Young's moduli, $E_0$ (\textbf{Fig. \ref{fig: stress-strain}.E}), following power-law scaling with respect to $\dot \epsilon$, $T_u \propto \dot{\epsilon}^{0.05}$ ($R^2 = 0.81$) and $E_0 \propto \dot{\varepsilon}^{0.11}$ ($R^2=0.97$), respectively. Meanwhile, raft extensibility - characterized by the strain, $\epsilon_u$, at which peak stress occurred - follows a weak inverse power-law scaling with respect to $\dot \epsilon$ (\textbf{Fig. \ref{fig: stress-strain}.F}, $\epsilon_u \propto \dot{\epsilon}^{-0.08}$, $R^2=0.66$).  Incidentally, decreased stiffness at lower strain rates offsets the slight gains in extensibility such that there is no statistically significant correlation between mechanical toughness, $u_t = \int T d \epsilon$, and the applied loading rate (see \textbf{Fig. \ref{SI: Delamination}.D} for toughness estimates).

    \figuremacro{H}{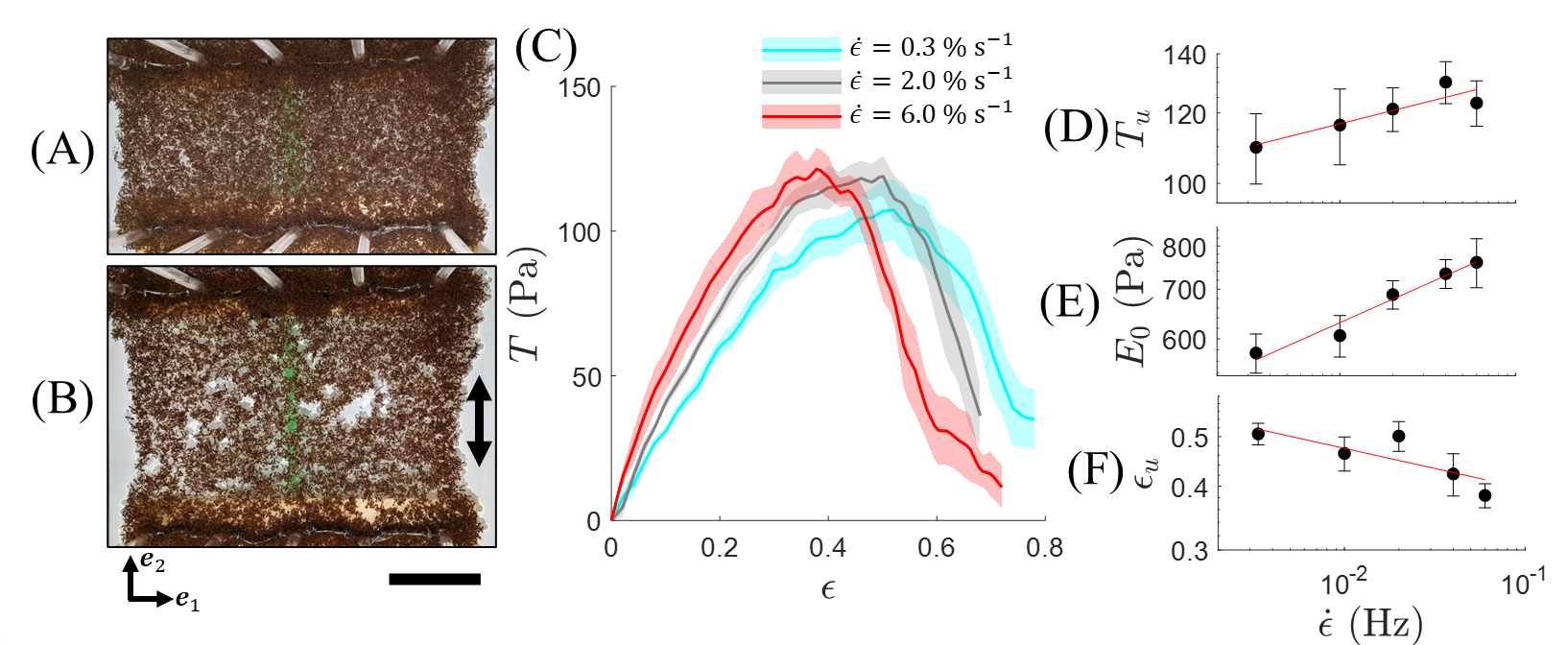}{}{\textbf{(A-B)} Top view of an ant raft undergoing uniaxial tension at \textbf{(A)} 0$\%$ and \textbf{(B)} 50$\%$ strain. The two-headed arrow in \textbf{(B)} denotes the loading direction. The scale bar represents 20 $\ell$ where 1 $\ell$ = 2.93 mm is the mean body length of an ant. Vectors $\bm e_1$ and $\bm e_2$ denote the orthonormal basis used throughout the remainder of this work, where $\bm e_2$ is the applied loading direction. \textbf{(C)} The ensemble average ($n=4$) of engineering stress, $T$, versus engineering strain, $\epsilon$, for ant rafts loaded at $\dot \epsilon=0.3\%$ Hz (cyan), $\dot \epsilon=2\%$ Hz (grey), and $\dot \epsilon=6\%$ Hz (red). See \textbf{Fig. \ref{SI: Extended stress}} for the extended raw stress data. Shaded regions represent standard error of the mean (S.E.). \textbf{(D-F)} ensemble-averaged ($n=4$) values of \textbf{(D)} Peak engineering stress, $T_u$, \textbf{(E)} small strain Young's modulus, $E_0$ (taken as the average tangent up to 0.1 strain), and \textbf{(F)} engineering strain, $\epsilon_u$, at peak stress are plotted with respect to $\dot \epsilon$. \label{fig: stress-strain}}{0.9}
    
This rate-dependent response is reminiscent of both biological \cite{desprat_creep_2005} and synthetic \cite{xu_thermosensitive_2022} dynamically bonded polymers, albeit far less pronounced. The rate-dependent behaviors of biological swarms are known to be driven by several characteristic times scales \cite{mora2016local}. The network reconfiguration time scale, $\tau_{net}$, (often taken as $1/k_d$ in the context of transient networks \cite{vernerey_statistically-based_2017,wagner_network_2021}) is associated with topological transitions (such as bond dissociation and neighbor exchange), while the conformational relaxation time scale, $\tau_{relax}$, deriving from strain localization and constituent re-alignment (without bond reformation) is governed by effects such as viscous drag, friction between neighboring constituents, and intrinsic relaxation in bonds. When dynamic networks are loaded at strain rates with characteristic times, $\tau_{strain} = 1/\dot \epsilon$, larger than $2 \tau_{net}$ (i.e., $W < 0.5$), their ability to reconfigure during loading often allows them to withstand strains on the order of 1,000$\%$ without damage \cite{cai_highly_2022} and increases in toughness on the order of 500$\%$ \cite{ducrot_toughening_2014,tong_highly_2019,xu_thermosensitive_2022,guo_understanding_2019,hubbard_creep_2022,li_role_2022}. In contrast, when $\tau_{strain} \leq 2 \tau_{net}$ (or $W \geq 0.5$), these networks' topologies remain relatively fixed enough to display the responses of time-dependent solids whose relaxation originates from strain localization and member realignment at characteristic time, $\tau_{relax}$ (rather than network restructuring). Our results suggest that under the prescribed loading condition, the ant rafts fall into the latter category, as they experience rupture at peak strains near 80$\%$ with no significant improvement in toughness at slower strain rates. Thus, rate-dependence is likely dominated by conformation changes. 
     
To corroborate the presence of and evaluate rate-dependent changes in the ants' local conformations (i.e., via strain localization, realignment, etc.) we conducted particle image velocimetry (PIV) which provides the instantaneous velocities, $\bm v(\bm x,t)$, of the structural raft layer as a function of position, $\bm x$, and time, $t$. See \textbf{Section \ref{sec: Methods: Damage-based Image Analysis}} for details of the analysis and mathematical definitions of the measures discussed here. \textbf{Fig. \ref{fig: PIV results}.A-B} illustrates snapshots of PIV results for a raft loaded at $0.3\%$ s$^{-1}$. For extended examples of PIV-gotten heat maps, see \textbf{Figs. \ref{SI: Velocity heat map}-\ref{SI: Curl heat map}}. To visualize local raft motion, the vector field denotes the average direction of local movement, $\bm {\hat \varphi} (\bm x)$ defined through Eqn. (\ref{eq: Normalized order parameter}). The heat map depicts the magnitude of deviation, $\delta \bm v^*$, between the \say{expected} velocity, $\bm v_{app}$, (based on applied loading conditions) and the measured velocity, $\bm v$, defined by Eqns. (\ref{eq: Applied velocity gradient}-\ref{eq: Velocity deviation}). Notably, this heat map highlights regions of quickly occurring localized strain, which occur predominantly adjacent to regions in which defects (i.e., voids) nucleate and coalesce as discussed further in \textbf{Section \ref{sec: Results: Void Nucleation}} and indicated by the onset of noise in \textbf{Figs. \ref{SI: Nonaffine speed}-\ref{SI: Raw spin}}. The transverse and axial components (normal to and in-line with the loading directions, respectively) of $\delta \bm v^*$, normalized by the applied loading speeds, are plotted with respect to the Weisenberg number in \textbf{Fig. \ref{fig: PIV results}.D} and \textbf{\ref{fig: PIV results}.E}, respectively. Here, the operator $\bar{\square}$ denotes spatiotemporal averaging over the entire raft and up to the time at which $30\%$ strain occurs (roughly the strain at which nucleated damage heavily confounds mechanical response in all cases - see \textbf{Section \ref{sec: Results: Void Nucleation}} and \textbf{Figs. \ref{SI: Nonaffine speed}-\ref{SI: Raw spin}}). We also examine the average rates of expansion (tr$(\bar{ \bm D}^*)>0$), contraction (tr$(\bar{\bm D}^*)<0$), counter-clockwise spin ($\bar \omega_{12}^*>0$), and clockwise spin ($\bar \omega_{12}^*<0$) (see \textbf{Figs. \ref{fig: PIV results}.E-I}, respectively), where $\bm D(\bm x,t)$ and $\bm \omega(\bm x,t)$ are the local rate of deformation and rate of spin tensors defined by Eqns. (\ref{eq: Rate of Deformation}) and (\ref{eq: Rate of Spin}), respectively. The asterisk after $\bm D$ and $\bm \omega$ denotes normalization by the applied strain rate, $\dot \epsilon$.

Prior to strain rate normalization, all localization measures are highly correlated with $\dot \epsilon$ (\textbf{Fig. \ref{SI: Non-normalized localization}}), simply suggesting that faster loading rates induce faster localization effects. However, normalizing $\delta \bm v$ by the loading speed reveals that average deviation speed (from the applied velocities) in the transverse (\textbf{Fig. \ref{fig: PIV results}.D}) and loading (\textbf{Fig. \ref{fig: PIV results}.E}) directions are on the order of $10-20\%$ and $15-25\%$ of the overall applied loading speed, respectively. Meanwhile, normalization of the expansion/contraction and spin rates by the applied strain rates reveals that the expansion (\textbf{Fig. \ref{fig: PIV results}.F}) and contraction (\textbf{Fig. \ref{fig: PIV results}.G}) are only on the order of $0.5-1\%$ of the overall applied strain rate, and the rates of spin are even smaller at $\sim 0.2-0.8\%$ of $\dot \epsilon$ (\textbf{Fig. \ref{fig: PIV results}.H-I}). 

    \figuremacro{H}{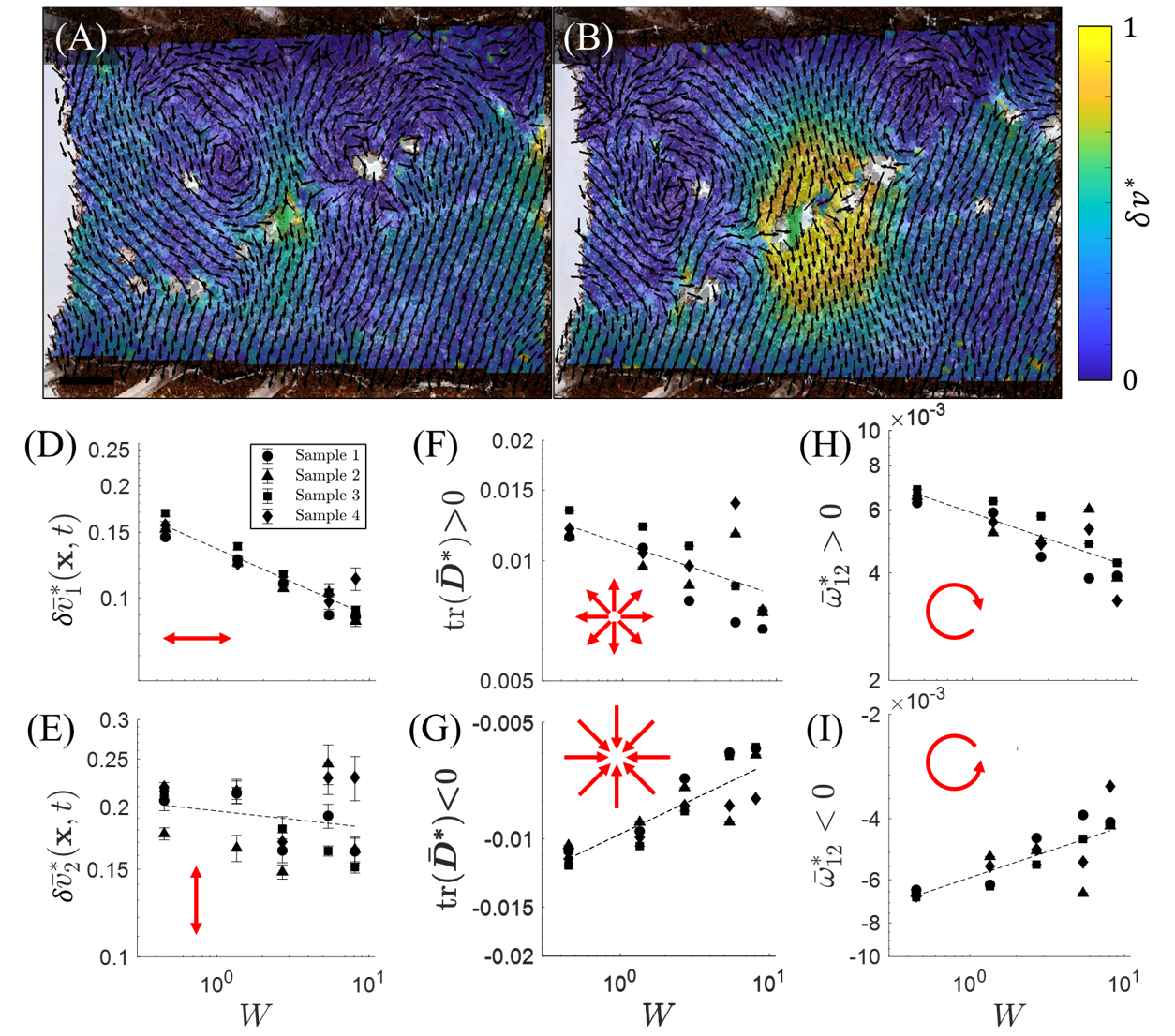}{}{ \textbf{(A-F)} Snapshots of a raft analyzed via PIV undergoing uniaxial tension at a Weissenberg number of $W=0.45$. See \textbf{Figs. \ref{SI: Velocity heat map}-\ref{SI: Curl heat map}} for samples heat maps of $v$, $\delta v^*$, tr$(\bm{D})$, and $\omega_{12}^*$, respectively. The vector field indicates the local, instantaneous direction of raft deformation averaged over a $[ 3 \times 3 ] \ell^2$ window, while the color map represents the degree of deviation, $\delta v^*$, from the expected/applied motion. The scale bar represents $10 \ell$ and the color bar is normalized by the applied loading speed at the displacement boundary. \textbf{(D-E)} The ensemble-averaged degrees of \textbf{(D)} transverse and \textbf{(E)} axial deviation from the applied velocity ($\delta \bar v_1^*$ and $\delta \bar v_2^*$, respectively) are plotted with respect to normalized strain rate (i.e., Weissenberg number, $W = \dot \epsilon / \bar k_d^0$) for $n=4$ samples. The red arrows indicate the axis of measured deviation for each plot. \textbf{(F-G)} The spatiotemporally averaged degrees of \textbf{(F)} expansion (tr$(\bar{\bm{D}}^*)>0$) and \textbf{(G)} contraction (tr$(\bar{\bm{D}}^*)<0$) are plotted with respect to $W$ for $n=4$ samples. The red arrows illustrate the senses of expansion and contraction (i.e., local sources and sinks, respectively). \textbf{(H-I)} The spatiotemporally averaged degrees of \textbf{(H)} counter-clockwise ($\bar{\omega}_{12}^*>0$), and \textbf{(I)} clockwise ($\bar{\omega}_{12}^*<0$) spin are plotted with respect to $W$ for $n=4$ samples. The red arrows illustrate the rotational directions. Note that while $\omega_{12}^*<0$ and $\omega_{12}^*>0$ are here plotted separately to illustrate their relative symmetry, their power-law scaling is reported for $[$det$(\bar{\bm \omega})]^{0.5}$ (since $\bar \omega_{12} \approx -\bar \omega_{21} \approx [$det$(\bar{\bm \omega})]^{0.5}$). \label{fig: PIV results}}{0.9}
    
Significantly, after normalization, all measures of the localization rate relate to $W$ following power-law scaling as $y\propto W^\nu$, where $y$ is a stand-in for any of the localization rate measures of \textbf{Fig. \ref{fig: PIV results}.D-I} (e.g., $\delta \bar v^*_1$) and $\nu$ characterizes the degree to which $y$ is disproportionate to $\dot \epsilon$ (whereby $\nu = 0$ indicates $y\propto \dot \epsilon$). Significantly, for all measures of localization degree except the velocity deviation in the loading direction, $\delta v_2^*$, the exponent, $\nu$, is statistically less than zero (and around negative $0.13-0.18$), indicating that rafts loaded at slower rates exhibit disproportionately greater degrees of localized transverse deformation ($\delta v^*_1 \propto W^{-0.18\pm 0.03}$, $R^2 = 0.90$, \textbf{Fig. \ref{fig: PIV results}.D}), expansion (tr$(\bm D^*) \propto W^{-0.13\pm 0.06}$, $R^2 = 0.78$, \textbf{Fig. \ref{fig: PIV results}.F}), contraction (tr$(\bm D^*) \propto W^{-0.18 \pm 0.04}$, $R^2 = 0.78$, \textbf{Fig. \ref{fig: PIV results}.G}), and spin ($[$det$(\bar{\bm \omega}^*)]^{0.5} \propto W^{-0.15\pm 0.05}$, $R^2 = 0.65$, \textbf{Fig. \ref{fig: PIV results}.H-I}) than rafts loaded at higher rates. These mechanisms of strain localization are tantamount to non-affine deformation modes \cite{picu_mechanics_2011,wagner_network_2021}, which allow network structures relax stress without necessarily having to restructure. Thus, they explain the modestly observed rate-dependence of \textbf{Fig. \ref{fig: stress-strain}.C-F}. Yet, they do not discount the presence of network restructuring and reveal nothing of the dissociative kinetics that induce void nucleation and eventual raft failure. Nor do they elucidate the constitutive mechanisms responsible for the rafts' brittle responses in the first place. 

\subsection{Damage signatures evince force-induced bond-stabilization}
\label{sec: Results: Void Nucleation}

The origins of brittle behavior and low failure strains, even when $W=0.45$, likely result from a couple of key differences between the ants and commonly investigated polymers. First, rather than resembling long polymer chains, ants have a small aspect ratio ($\sim 4:1$ \cite{tschinkel_morphometry_2013}) so that they are not prone to the uncoiling and reptative disentanglement \cite{rubinstein_polymer_nodate} that bestow polymers with high extensibilities. Second, bond exchange-imparted creep in polymers is facilitated by their near incompressibility, which conserves the concentration of open binding sites and therefore the rate of bond attachment \cite{stukalin_self-healing_2013,wagner_coupled_2023}. This stems from polymer chains' cohesion not only by crosslinks, but also inter-chain Van der Waals interactions, electrostatic potentials, and hydrogen bonding \cite{brandt_calculation_2004}. Recently, Ko \textit{et al.} (2022)\cite{ko_small_2022} have demonstrated that in addition to ant-to-ant bonds, floating ants are weakly cohered by surface tension - a phenomenon dubbed the \say{Cheerios effect}. However, the resulting cohesion is not strong enough to sustain near-incompressibility in rafts of the size studied here. Consequently, in the presence of biaxial stress, as is the case for our experimental conditions, the rafts sustain increases in their surface areas, causing defects to occur in distinct phases of nucleation/growth and coalescence, as illustrated through \textbf{Fig. \ref{fig: void coalescence}}. 
    
Towards understanding the dissociative mechanisms driving void nucleation and coalescence, we here characterize rafts' topological signatures using image analysis. See \textbf{Figs. \ref{fig: damage-analysis}.A-B} or \textbf{\ref{SI: Uniaxial image analysis}} for annotated depictions of the analysis. Rafts' free volume fractions, $\phi_v$, (i.e., the areal fraction visually unoccupied by ants), minimum cross-sectional lengths, $L_x$, normal to loading (i.e., in direction $\bm e_1$), average void areas, $A_v$, and number of voids, $N_v$ were computed using the methods detailed in \textbf{Section \ref{sec: Methods: Damage-based Image Analysis}}. Voids are defined as continuous regions absent of ants whose size exceeds the approximate occupancy or \say{reach} area, $A_r \approx \pi \ell^2$, for an ant of body length $\ell$. \textbf{Figs. \ref{fig: damage-analysis}.A} displays an unaltered and image-processed snapshot of an ant raft undergoing uniaxial tension at $\epsilon = 50\%$. The dashed red lines, blue dots, and solid red lines in \textbf{Fig. \ref{fig: damage-analysis}.A} denote the analyzed region of interest, centroids of defects, and cross-section with the minimum length, $L_x$, respectively.  

    \figuremacro{H}{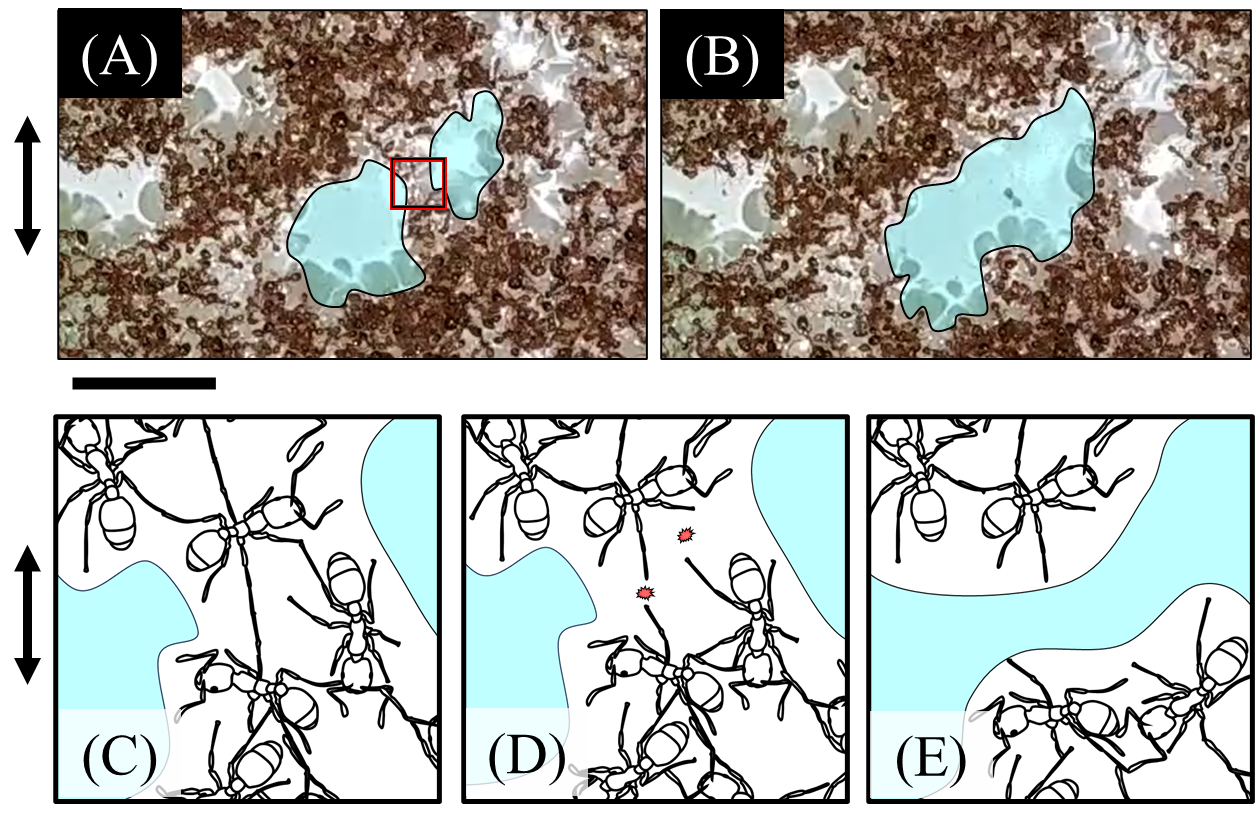}{}{\textbf{(A-B)} Close ups of a raft section illustrate a void coalescence event in which two voids (highlighted in blue) merge due to the strain-driven dissociation of an ant filament, which was originally separating the voids. The duration between \textbf{(A)} and \textbf{(B)} is one second and the scale bar represents 5$\ell$. \textbf{(C-E)} A zoomed-in schematic, representing the network's structure outlined in red in \textbf{(A)}, illustrates the mechanism of void coalescence with voids being highlighted blue. The schematic depicts a local filament of networked ants that bridge a gap and partition two flanking voids \textbf{(C)} while the filament is intact, \textbf{(D)} immediately after the ants comprising it dissociate (denoted by red stars), and \textbf{(E)} after void coalescence. \textbf{(A-E)} The two-headed arrows are aligned with the direction of applied loading. \label{fig: void coalescence}}{0.6}
    
Neither the ensemble-averaged ($n=4$) change in free volume fraction, $\phi_v$, (from the initial value) nor the minimum cross sectional length, $L_x^*$ (normalized by rafts' initial widths) vary discernibly with respect to strain rate, $\dot \epsilon$ (\textbf{Figs. \ref{SI: Extended ensemble average damage}}). Likewise, neither the ensemble-averaged void area, $A_v^*$ (normalized by $A_r$), nor number of voids, $N_v^*$, appear to vary reliably with loading rate (\textbf{Fig.\ref{SI: Extended ensemble average damage}.E-H}). Mean values of void area, $A_v$, are consistently between 5 and 10 times the occupancy area, $A_r$, regardless of $\dot \epsilon$ (\textbf{Figs. \ref{SI: Extended ensemble average damage}.E-F}). Meanwhile, the numbers of voids, $N_v$, are consistently between 2 and 4 defects per $10^3$ ants (\textbf{Figs. \ref{SI: Extended ensemble average damage}.G-H}), regardless of loading rate, highlighting the relative sparsity of voids needed to induce coalescence and failure. While the number of voids, $N_v^*$, appears to increase somewhat steadily with respect to strain (\textbf{Fig. \ref{fig: damage-analysis}.C} and \textbf{Figs. \ref{SI: Extended ensemble average damage}.G-H}), void areas, $A_v^*$, tend to increase drastically once applied strain exceeds 40$\%$ (\textbf{Fig. \ref{fig: damage-analysis}.D} and \textbf{Figs. \ref{SI: Extended ensemble average damage}.E-F}). The sudden increase in $A_v^*$ is indicative of visible merging between neighboring defects as the load paths distinguishing them fracture (e.g., the filament depicted in \textbf{Fig. \ref{fig: void coalescence}}).

Since defect nucleation (indicated by increasing $N_v^*$) evidently precedes coalescence (indicated by increasing $A_v^*$), a characteristic lag strain, $\epsilon_c$, between these two phenomena may be measured (\textbf{Fig. \ref{fig: damage-analysis}.C-D}) and used to define the corresponding lag time, $\tau$. This lag provides insight to the time- and stretch-dependence of local dissociation events. Having observed that void coalescence is driven by the failure of inter-void ligaments (e.g., \textbf{Fig. \ref{fig: void coalescence}}), then $\epsilon_c$ may either depend on (a) the strain-dependent (and time-independent) rupture of ant-to-ant connections due to over-stretching, (b) the time-dependent (and strain-independent) stochastic detachment of these connections (at rate $\bar k_d^0$), or (c) some combination of these mechanisms. If the dissociative lag depends only on bond stretch, then the lag strain $\epsilon_c$ should be roughly constant, regardless of loading rate, so that $\tau \propto \dot \epsilon^{-1}$ (\textbf{Fig. \ref{fig: damage-analysis}.E}, dashed blue line). On the other hand, if the lag strain depends only on the elapsed time because it is governed entirely by stochastic, strain-independent bond dissociation, $\tau$ should remain constant so that $\tau \propto \dot \epsilon^0$ (\textbf{Fig. \ref{fig: damage-analysis}.E}, dashed green line). The experimental lag strain was computed as the strain at which peak cross-correlation between $N_v^*(\epsilon)$ and $A_v^*(\epsilon)$ occurs per Eqn. (\ref{eq: Cross-correlation function}). See \textbf{Section \ref{sec: Methods: Damage-based Image Analysis}} for details of analysis and \textbf{Fig. \ref{SI: Cross-correlation}} for a sample of the cross-correlation function for the extreme loading rates investigated. The resulting experimental data (\textbf{Fig. \ref{fig: damage-analysis}.E}) reveals that $\tau \propto \dot \epsilon^{-0.75}$ ($R^2=0.90$) suggesting that the void coalescence state is mostly mediated by strain (i.e., bond stretch), rather than elapsed time (i.e., stochastic dissociation at the rate $\bar k_d^0$). Thus, this indicates that the merging of voids due to bond fracture is primarily driven by the finite extensibility of connections. This is consistent with the findings of Phonekeo, \textit{et al.} (2017) \cite{phonekeo_ant_2016} who vertically loaded 3D assemblages of fire ants and reproduced their experimental observations with a dissociating lattice model presuming force-induced stabilization or \say{catch bond} kinetics \cite{sokurenko_catch_2008}. 
    
Catch bond kinetics impose that dynamic bonds remain attached for longer, probabilistic lifetimes when under some range of applied load than at rest. The hypothesized presence of catch bond kinetics may explain why - even when $W=0.45$ - the ants exhibit brittle, elastic behavior rather than viscoelastic creep. The nominal dissociation rate, $\bar k^0_d \approx 7.4$ mHz, was measured in unperturbed rafts subject only to internally generated contractile forces (\textbf{Fig. \ref{SI: Active contraction}}) on the order of $10^{-2}$ N, rather than externally applied loading, which generated forces four orders of magnitude higher ($\sim 10^2$ N) for comparably sized rafts. It seems likely that $k_d$ decreases for mechanically loaded ant bonds, so that the effective Weissenberg numbers are higher than the nominally estimated values. Therefore, rather than dissociate approximately every $\bar k_d^{-1} \approx 140$ seconds, ant-to-ant bonds may instead dissociate predominantly when extended to some finite strain at which force diverges due to stretching (rather than unfolding) of the legs, akin to polymers stretched near their contour lengths \cite{lamont_rate-dependent_2021,mulderrig_statistical_2023}. 

    \figuremacro{H}{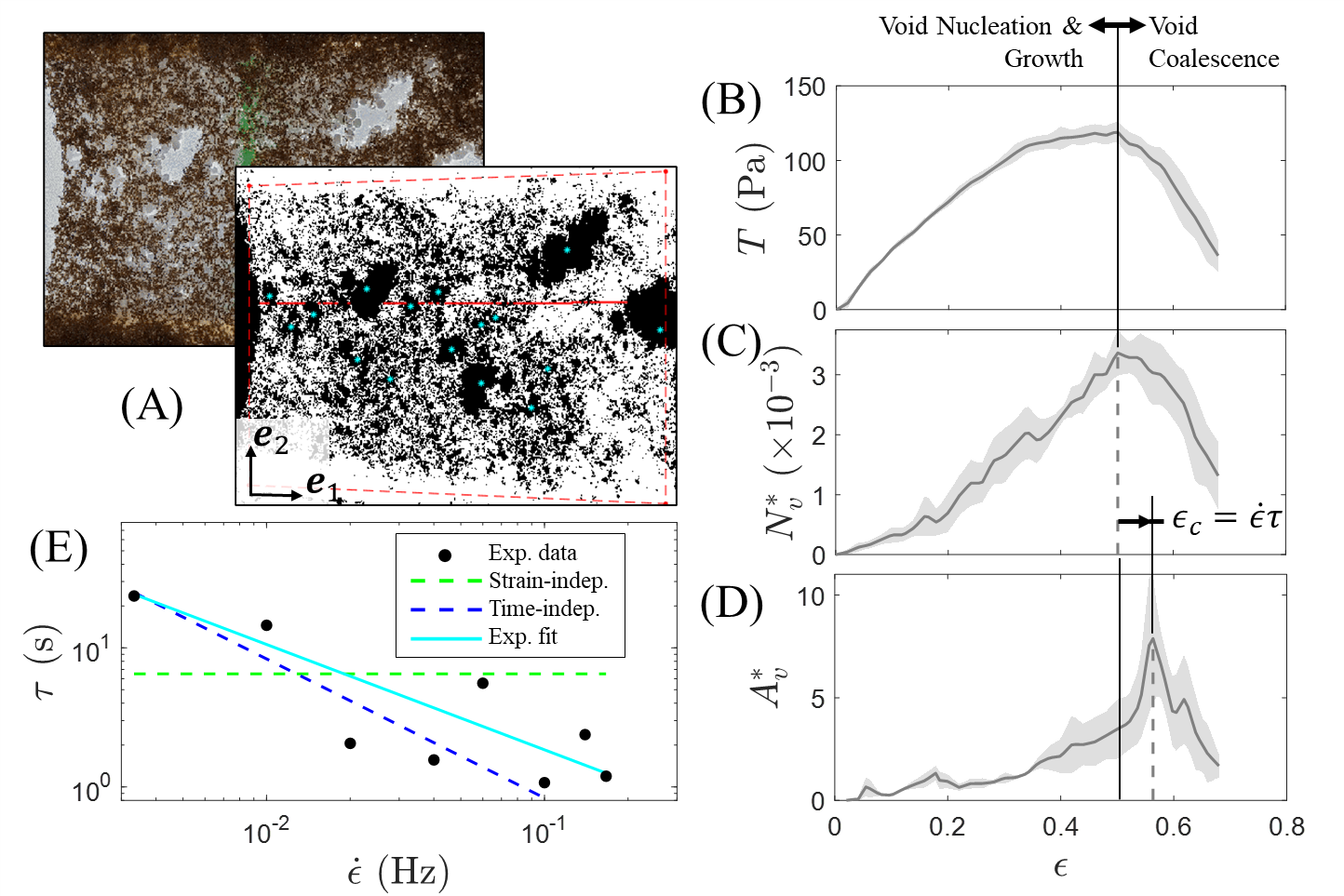}{}{\textbf{(A)} An unprocessed and annotated frame (top view) of an ant raft undergoing uniaxial tension at $\sim 50\%$ strain. Dotted red lines in the annotated image denote the boundaries of the analyzed region of interest, blue dots denote voids' centroids, and the solid red line transversing the raft highlights the minimum cross-sectional length, $L_x^*$. The scale bar represents $10 \ell$. \textbf{(B-D)} The ensemble-averaged ($n=4$) \textbf{(B)} stress, \textbf{(C)} void count, and \textbf{(D)} void area are plotted with respect to strain for the intermediate experimental loading rate, $\dot \epsilon = 2\%$ s$^{-1}$. See \textbf{Fig. \ref{SI: Extended ensemble average damage}} and \textbf{\ref{SI: Extended raw damage}} for the extended ensemble-averaged and raw damage analysis data, respectively. Critical lag strain, $\epsilon_c$, is defined as the strain that elapses between when the peak number, $N_v^*$, and area, $A_v^*$, of voids occur. Shaded regions represent S.E. \textbf{(E)} Experimentally estimated failure lag time, $\tau = \epsilon_c/\dot \epsilon$, (black circles) is plotted with respect to loading rate, $\dot \epsilon$. Lines depict power-law scaling fits of the form $\tau = \tau_0 (\dot \epsilon/\dot \epsilon_0)^{-\nu}$, where $\tau_0$ is simply the lag time when $\dot \epsilon = \dot \epsilon_0$ and $\dot \epsilon_0$ is arbitrarily fixed at 1 Hz. The dashed green and blue lines represent the extreme ideal cases that $\tau$ is a constant regardless of elapsed strain (i.e., $\delta \epsilon \propto \dot \epsilon$ so that $\nu=0$, $R^2 = 0.00$) and $\delta \epsilon$ is a constant regardless of elapsed time (i.e., $\tau \propto \dot \epsilon^{-1}$ so that $\nu=1$, $R^2 = 0.86$), respectively. The solid cyan line is the power-law fit to experimental data, ($\tau_0 = 1.93$ s, $\dot \epsilon_0 = 0.10$ Hz, $\nu = -0.75$, $R^2=0.90$). \label{fig: damage-analysis}}{0.9}

\subsection{Rafts self-heal upon reintroduction of initial densities}

While the conditions of dissociative kinetics are investigated by the damage analysis of \textbf{Section \ref{sec: Results: Void Nucleation}}, the associative kinetics remain unexplored. We previously supposed that the presence of defects is detrimental to the rafts' mechanical recovery as ants are unable to reconnect across large voids, which retards the associative kinetics intrinsic that we hypothesize would otherwise enable steady-state creep or even self-healing (\textbf{Fig. \ref{fig: void coalescence}}). To gauge this supposition and probe the associative kinetic timescale, we cyclically loaded rafts twice to $50\%$ strain with hold times of $t_h=\{ 0,10,30,60,300 \}$s between cycles. To probe effects of rate-dependence, we conducted these test at $\dot \epsilon =\{ 2,4 \} \%$ s$^{-1}$. Engineering stress-strain results of these experiments are depicted in \textbf{Fig. \ref{cyclic stress-strain}.A-B}. Isolating the mechanical response for the extreme hold times, $t_h = \{0,300 \}$s, it becomes clear that for both strain rates, the longer hold time of 300 s permitted better mechanical recovery, suggesting a greater degree of restorative crosslinking, as expected. 
   
    \figuremacro{H}{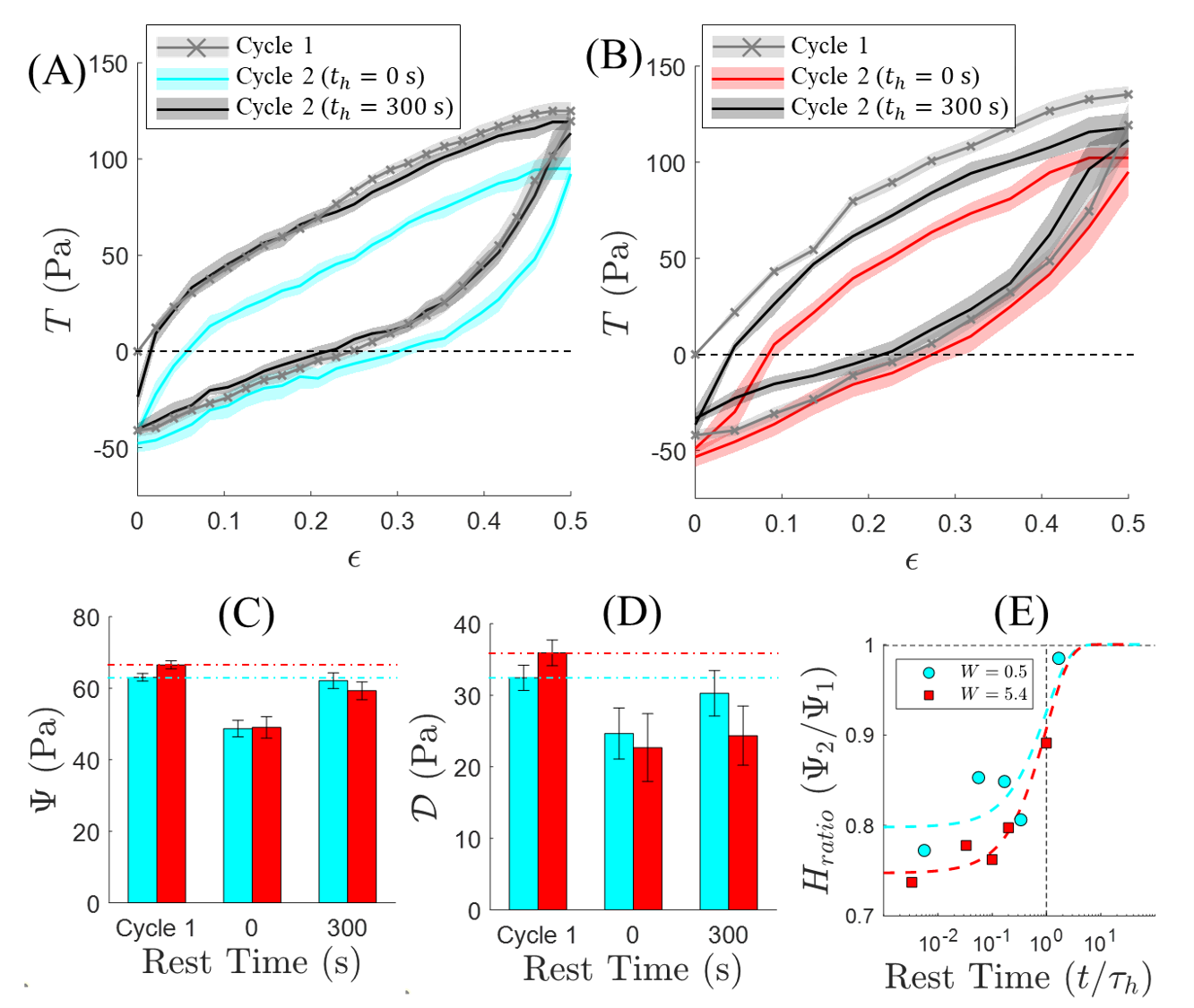}{}{\textbf{(A-B)} ensemble-averaged ($n=4$) engineering stress versus strain (to $50\%$) over two loading cycles with recovery times of $t_h \in \{ 0, 300 \}$s for \textbf{(A)} $W=2.7$ and \textbf{(B)} $W=5.4$. See \textbf{Fig. \ref{SI: ensemble-averaged cyclic stress}} for extended data. The first loading cycle is ensemble-averaged for all $n=20$ experiments (grey curve with $\times$-shaped markers), while the second load cycles are depicted as solid curves. \textbf{(A-B)} Shaded regions indicate S.E. \textbf{(C-D)} Ensemble-averaged \textbf{(C)} strain energy, $\Psi$, and \textbf{(D)} dissipated energy, $\mathcal{D}$, for each of the five recovery times when $W=2.7$ (cyan bars) and $W=5.4$ (red bars). Error bars represent S.E. See \textbf{Fig. \ref{SI: Extended strain energy}} for the total strain energy, $\Psi$, stored energy (during unloading), $\Psi_{el}$, and dissipated energy, $\mathcal{D}$, at all hold times. \textbf{(E)} Hysteresis ratio with respect to recovery time, $t/\tau_h$, from experiments (discrete data) and an analytical fit (dashed curves). \label{cyclic stress-strain}}{0.75}
    
To quantify mechanical recovery across all hold times total strain energy, $\Psi = \int_0^{0.5} T \partial \epsilon$ (integrated over the loading curves), elastically stored energy, $\Psi_{el} = \int_0^{0.5} T \partial \epsilon$ (integrated over the unloading curves), and dissipated energy $\mathcal{D}=\Psi - \Psi_{el}$, were also computed. Total, $\Psi$, and dissipated, $\mathcal{D}$, energies are reported in \textbf{Fig. \ref{cyclic stress-strain}.C-D} for both cycles. Results are combined for the ensemble average ($n=20$) of all first loading cycles (since they shared identical conditions), but provided separately for the second loading cycles at the various hold times, $t_h$ ($n=4$). There is no statistically significant difference between the total, $\Psi$, or dissipated, $\mathcal{D}$, energy between the first and second cycles loaded at $2\%$ s$^{-1}$ and held for $300$ s between cycles, indicating complete mechanical recovery and restoration of bond concentration. However, for smaller hold times, there is a statistical reduction in all three measures (\textbf{Fig. \ref{cyclic stress-strain}.C-D} and \textbf{\ref{SI: Extended strain energy}}), indicative of softening between the first and second cycle. Greater restoration of mechanical properties at longer hold times provides strong evidence that attachment rate, and therefore transient crosslink concentration, increases when raft densities are restored to their initially high values.
    
Regarding the effects of loading rate, \textbf{Fig. \ref{cyclic stress-strain}.C} reveals that the average strain energy of the first cycle for rafts loaded at $2\%$ s$^{-1}$ is statistically lower than that for rafts loaded at $4\%$ s$^{-1}$, which is consistent with the strain rate-driven stiffening demonstrated by \textbf{Fig. \ref{fig: stress-strain}.A}. Despite this, there is no statistically significant difference in stored energy, $\Psi_{el}$, (\textbf{Fig. \ref{SI: Extended strain energy}}) for the first (or second) cycles across the two strain rates, confirming along with \textbf{Fig. \ref{cyclic stress-strain}.D} that more energy is dissipated by the rafts loaded at faster rates. That stored energy, $\Psi_{el}$, is indistinguishable for the two loading rates suggests that very similar network structures (i.e., crosslink densities) exist in the rafts after loading to $50\%$ strain. This is consistent with the finding of indistinguishable damage characteristics (\textbf{Fig. \ref{SI: Extended ensemble average damage}}) across loading rates, and supports the notion that dissipative dissociation kinetics are driven by finite bond extensibilities. Nevertheless, complete recovery is observed in rafts loaded at the slower rate when $t_h = 300$ seconds, but not in rafts loaded at the faster rate and held for the same amount of time.
    
To extrapolate predictions about bond kinetics from this data, we examine the hysteresis ratio, defined as $H_{ratio}=\Psi_2/\Psi_1$, where $\Psi_1$ and $\Psi_2$ are the total strain energies of the first and second cycles, respectively. From this ratio, we may predict a characteristic heal time, $\tau_h$, according to:
\begin{equation}
    H_{ratio} = 1 - \Delta \Psi^* \exp (-t_h/\tau_H)  
    \label{H_ratio}
\end{equation}
where $\Delta \Psi^*$ is the fractional reduction in strain energy from cycles one to two when recovery time is $t_h = 0$, and $\tau_h$ is defined as the recovery time required for a network to regain 63$\%$ of its lost strain-energy carrying capacity \cite{lamont_rate-dependent_2021}. Plotting $H_{ratio}$ with respect to $t_h$ (\textbf{Fig. \ref{cyclic stress-strain}.H}) and fitting (\ref{H_ratio}) to the data provides that the characteristic heal time of the rafts loaded at $2$ and $4\%$ s$^{-1}$ are approximately $180$ and $300$, respectively. Lamont, \textit{et al.} (2021) \cite{lamont_rate-dependent_2021} have demonstrated analytically that $\tau_h$ for dynamic networks is related to both dissociative and associative bond kinetics, scaling as $\tau_h = (\bar k_a+\bar k_d)^{-1}$. This suggests that the higher loading rate induces a reduction in $\bar k_a$, $\bar k_d$, or both and that this effect persists during the hold times between cycles. Supposing that $k_a$ depends only on the local concentration of ants (so that it is constant during all recovery stages), then this may provide secondary evidence of catch bond kinetics, whereby $\bar k_d$ increases when higher loading rates are applied. Unlike polymers, ants have behavioral agency and may change their preferential bonding characteristics based on their stimulus history. For example, a behaviorally driven reduction of $\bar k_d$ during loading (especially at higher strain rates) could temporarily reduce the exchange rate that promotes healing during subsequent recovery times.

\section{Conclusion}
We here loaded planar fire ant rafts under uniaxial extension at variable loading rates. Generally, the rafts are highly compliant structures, with network-scale peak stresses on the order of $10^2$ Pa. While we observed measurable, albeit small, strain rate-dependence for peak stresses and initial tangent moduli, we found no significant correlation between mechanical toughness or rupture strains and the investigated strain rates. Significantly, despite rate-dependence, brittle-like behavior persisted even in the slow loading regime where Transient Network Theory predicts steady-state creep onset, based on the measured rate of unperturbed ant-to-ant dissociation. This brittle-like response is attributed to a combination of force-induced bond stabilization, and reduction in bond association kinetics due to biaxial stress-driven decrease in the planar ant concentration (and therefore a smaller number of ant-to-ant bond opportunities). Meanwhile, the small degree of  rate-dependence observed is attributed to measurable differences in conformational relaxation without major network restructuring. Specifically, PIV analysis revealed that rafts loaded at slower rates exhibited disproportionate degrees of localized transverse deformation, expansion, contraction, and spin as compared to rafts loaded at faster rates. Notably, conformational rate-dependence exists even in the strain regime absent of widespread bond scission, damage onset, or other major raft restructuring (i.e., $0\leq \epsilon \leq 0.3$).   

To elucidate any rate-dependence in dissociation kinetics, we analyzed the damage characteristics of rafts via custom image analysis. We found that lag strain between the peak number and area of voids is predominantly governed by the stretch of the network, rather than the duration of elapsed time. This implies that the underlying bond rupture events driving void coalescence depend more on finite bond extensibility than voluntary ant detachment events, providing further evidence force-stabilizing bonds. Nevertheless, stochastic dissociation events should not be discounted and it is possible that loading rafts at even slower rates would allow bond exchange-enabled viscous creep. However, in our previous work, we observed that ants exit the structural layer and enter the freely active top layer at a rate of $\delta \approx 0.33$ mHz \cite{wagner_treadmilling_2021}. Given the boundary conditions applied here, we find that this culminates in dissolution of the ant rafts over hour-long timescales ($1/\delta = 1/0.33$ mHz$^{-1} \approx 0.84$ hrs) as if by some effective surface tension (\textbf{Fig. \ref{SI: Active contraction}}). This rendered mechanical loading of cohesive rafts at rates slower than $2\%$ s$^{-1}$ difficult. However, in future studies we may conduct experiments on rafts with a higher length-to-width ratio in order to reduce biaxial stress states, and therefore mitigate loss in planar ant concentration and bond association.

To corroborate that planar ant concentration modulates bond association kinetics (because of the finite reach of ants), we loaded rafts uniaxially to 50$\%$ strain for two consecutive cycles and with various hold times between cycles. Significantly, rafts demonstrate restoration of mechanical properties between the first and second loading cycle, verifying their concentration-dependent self-healing. Furthermore, rafts exhibited significantly better mechanical recovery for the longest hold times ($t_h=300$ s) than without recovery ($t_h=0$ s). Utilizing the ratio between strain energies of the second and first loading cycles, we extrapolated characteristic heal times related to both the attached and detached lifetimes of bonds during recovery. We found that while there is no statistically significant difference in the stored mechanical energy of rafts loaded at slower versus faster rates (confirming that rate-dependent mechanical response does not derive from rate-dependent network reconfiguration for the prescribed loading rates), there was a sizeable increase in heal time (180 s versus 300 s) for the rafts loaded at faster rates. This not only supports the notion that fire ants decrease their characteristic bond exchange times (implicating catch bonds), but also demonstrates that their altered bond kinetics persist even during the recovery time (after mechanical stress is removed). This perhaps showcases ants' capacity for complex mechanical memory as active, living constituents, which we will explore further in future work.    

\section{Methods}
\label{sec: Methods}
  
\subsection{Experimental setup}
\label{sec: Methods: Experimental Setup}

\textbf{Fig. \ref{SI: Exp setup}} depicts a model of the apparatus used to conduct tensile testing on ant rafts. The entire apparatus was suspended above a container of water such that only the bases of the substrates to which the ants adhered (Substrates 1 and 2), as well as the base of a tertiary substrate used to estimate force were submerged below the water line. Tensile strain was applied at a controllable engineering strain rate using a belt-driven linear actuator. The top of Substrate 1 was affixed to the moving plate of the linear actuator such that the entirety of Substrate 1 translated as the linear actuator was driven. A T-shaped wooden block was rigidly affixed to one end of the linear actuator. Substrate 2 was suspended from this block via three hinges that allowed it to pivot slightly. The tops of Substrates 1 and 2 were rigidly attached to their submerged bases via five acrylic rods. These rods were each coated with PTFE and talcum powder to prevent the ants from climbing the apparatus and escaping the water. 

At the base of Substrates 1 and 2, Velcro was wrapped around all five rods such that the ants could easily envelop and adhere it. Thus, the ants could form a continuous raft suspended between the two substrates and as Substrate 1 translated, the ants underwent uniaxial tension. To estimate force during loading, a spring was suspended underwater between the backside of Substrate 2, and the wooden bock fixed to the linear actuator track (Substrate 3). Top views of the ant rafts (\textbf{Fig. \ref{SI: Exp setup}.C}) were captured during loading via a camera positioned as shown in \textbf{Fig. \ref{SI: Exp setup}.B}. This footage was synched up with side-view footage (\textbf{Fig. \ref{SI: Exp setup}.D}) of the underwater spring. Force carried by the ants was estimated by visually measuring the horizontal component of spring stretch (see \textbf{Section \ref{sec: Methods: Force Estimation}} and \textbf{Fig. S2}). Notably, the spring’s elongation (and thus displacement of Substrate 2) was on the order of 10$^{-4}$-10$^{-3}$ m, while the overall raft displacement was on the order of 10$^{-1}$ m such that spring deflection had negligible effect on the overall raft strain during loading. Similarly, the angle of deflection of Substrate 2 was on the order of just 1$^\circ$ such that the change in z-axis position of Substrate 2 (and the ants attached to it) was less than 10$^{-4}$ m. 

\subsection{Ant collection and mechanical testing}
\label{sec: Methods: Ant Collection and Mechanical Testing}

Fire ant workers were collected from seven separate colonies in Keller, TX. Samples were prepared from one of the randomly selected colonies by slowly flooding the container in which it was housed until the ants surfaced for collection. Once isolated from debris, the ants were massed in a tared petri dish and allowed to rest for 30 minutes before being placed into the water between the Substrates 1 and 2 of the apparatus (\textbf{Fig. \ref{SI: Exp setup}}). Sample masses were consistently on the order of $10$ g, indicating that there were $\sim10^4$ ants in each raft based on an estimated ant mass of 1 mg.\cite{mlot_fire_2011,wagner_treadmilling_2021,tschinkel_morphometry_2013} Sample shape was achieved by placing the ants in the water between Substrates 1 and 2 of the apparatus, and then gently coaxing the rafts towards each substrate before gently cropping the excess raft material from the sides using forceps. Once in the correct shape, the rafts were allowed to rest for 10-minutes prior to testing. To ensure that differences in results were not due to colony selection, once a raft sample was set up, experiments were carried out at all reported strain rates (and/or hold times in the case of cyclic loading) before returning the ants to their habitats. To mitigate and control for any effects of cumulative ant exhaustion, the rafts were allowed to rest and self-heal at least 5-minutes between experiments. Additionally, the order of applied strain rates and hold times (for cyclic loading) was randomly assigned for each testing batch. No sample of ants was tested for more than 10 experiments before being returned to their habitat for overnight recovery. 

\subsection{Force estimation}
\label{sec: Methods: Force Estimation}

Forces transmitted by the ant rafts were estimated by measuring the strain of an elastomeric spring (\textbf{Fig. \ref{SI: Exp setup}.D}) of known force-strain relation in wet conditions (\textbf{Fig. \ref{SI: Force calibration}}). Strain was taken as $\epsilon=(|\bm{r}|-|\bm{r}_0 |)/|\bm{r}_0 |$ where $|\bm{r}|$ and $|\bm{r}_0 |$ are the current and reference (force-free) lengths of the spring, respectively, both of which were measured using automated image-analysis in MATLAB R2022b. While the spring was angled slightly out of the horizontal plane (by angle $\theta$), only the horizontal component of tension, $\epsilon_1=\epsilon \cos{\theta}$  resisting elongation of the ant raft was used to estimate resistance force and stress.

\subsection{Estimating unperturbed bond dynamic rates}
\label{sec: Methods: Estimating Unperturbed Bond Dynamic Rates}

To estimate the nominal bond dissociation, $\bar k_d^0$, and association, $\bar k_a^0$, rates of ant-to-ant bonds, top view footage of an unperturbed ant raft was used to individually measure the attached, $\tau_a$, and detached, $\tau_b$, bond lifetimes of image-tracked legs. Plotting the probability distribution functions of $\tau_a$ and $\tau_b$, we find that they each follow a Poisson’s process \cite{wagner_network_2021} and that the probabilities of finding a leg at a given attached or detached lifetimes of $\tau_a$ or $\tau_b$ respectively evolve according to $P_d = \exp (-\bar k_d^0 \tau_a )$ and $P_a = k_a \exp (-\bar k_a^0 \tau_d)$ ({\textbf{Fig. \ref{SI: Bond dynamics}}). Notably, mandible-enabled bonding modes were not commonly observed as in the case of 3D aggregations \cite{foster_fire_2014}. 
  
\subsection{Strain rate selection}
\label{sec: Methods: Strain Rate Selection}

Although rafts strained at rates on the order of $0.3$\% s$^{-1}$ could be reasonably tested, we encountered that rafts loaded at rates slower than this (e.g., $\sim0.01-0.1\%$ s$^{-1}$) were often subject to dissolution caused by the structural layer's perpetual contraction \cite{wagner_treadmilling_2021} and ants' preferential aggregation on the dry substrates rather than back into the structural raft network (\textbf{Fig. \ref{SI: Active contraction}}). Meanwhile, systems loaded at rates greater than $6\%$ s$^{-1}$ often delaminated at the ant-to-substrate interface, likely due to a combination of viscous drag from the water's surface and sudden onset of high local stress due to geometric stress risers at the junction. Either way, this delamination obfuscated trends in rate-dependent response (see \textbf{Fig. \ref{SI: Delamination}} for extended mechanical response data) at measured loading rates $>6\%$ s$^{-1}$ and so stress data from such rates is not reported in \textbf{Fig. \ref{fig: stress-strain}}. Note that while drag may have induced delamination, it is ruled out as a major source of uncertainty for the reasons discussed in \textbf{Section \ref{sec: SI: Drag}}. 

\subsection{Conformational PIV Image Analysis}
\label{sec: Methods: Conformational PIV Image Analysis}

To characterize any rate-dependent conformational changes in the ant rafts, PIV was utilized to obtain Eulerian vector fields of the structural ant raft during deformation using the application, PIVlab \cite{thielicke_particle_2021}, in MATLAB 2022b. Image stacks were extracted from raw footage such that the applied elongation between frames was 1 mm (roughly $\ell/3$), regardless of $\dot \epsilon$. This enabled a consistency of PIV settings across strain rates. To attain sufficient spatial sampling, the interrogation length was set to $\sim\ell/3$. A dynamic mask was applied to exclude voids (defined as regions where the vacant area exceeded areal reach, $A_r = \pi \ell^2$) from the regions of interest as they nucleated. The result was a spatiotemporally resolved instantaneous velocity field, $\bm v$ at all positions $\bm x$ on the structural raft (see \textbf{Fig. \ref{SI: Velocity heat map}} for examples). To reduce visual noise and highlight regions where ants in the the raft are moving more or less synchronously with their immediate neighbors, the normalized order parameter was computed as a function of position, $\bm x$, according to:
\begin{equation}
    \bm{\varphi} (\bm x) =\frac{\int_\Omega \bm v(\bm x)}{\int_\Omega  |\bm v (\bm x)|},
    \label{eq: Normalized order parameter}
\end{equation}
where $|\bm v|$ is local raft speed and the integral is carried out over a moving areal window of $\Omega = 3 \ell \times 3 \ell$ (or $\sim 10 \% \times 10 \%$ of the initial sample length, $L_0 \approx 34 \ell$). The order parameter is unity wherever the ants are moving unidirectionally within the domain, $\Omega$, and approaches zero when the motion is completely isotropic and disordered. The norm of $\varphi$ is shown as a vector field in \textbf{Fig. \ref{fig: PIV results}.A-B} and \textbf{Figs. \ref{SI: Velocity heat map}-\ref{SI: Curl heat map}} to illustrate the direction of local movement.

Ultimately, we wish to determine the local displacement everywhere on the raft with the intention to characterize the extent of local deviation in the raft from expected or \say{affine} behavior \cite{picu_mechanics_2011}. First, we consider the velocity field induced by the applied elongation. The velocity gradient applied at the boundaries is:
\begin{equation}
    \bm{L}_{app} = \text{diag}(0,\dot{\varepsilon}(t))
    \label{eq: Applied velocity gradient}
\end{equation}
where $\dot{\varepsilon}$ is the applied strain rate  in direction $\bm e_2$. Due to the non-normal camera perspective, $\dot \varepsilon$ was computed along the width of each sample (i.e., along direction $\bm e_1$), using a discrete difference approximation \cite{wagner_treadmilling_2021}:
\begin{equation}
    \dot{\varepsilon} \approx x_2(t)^{-1} \left[ x_2(t+\Delta t) -  x_2(t) \right]/\Delta t,
    \label{eq: Estiamted Strain Rate}
\end{equation}
where $x_2$ is the position of each point (in direction $\bm e_2$) along the moving boundary of the raft's region of interest and $\Delta t$ is the time between adjacent frames. Had the network followed $\bm L_{app}$ affinely, then the \say{applied} velocity could be taken as the linearly interpolated velocity between the two boundaries. This interpolation was conducted using $\bm{L}_{app}$ as follows:
\begin{equation}
     \bm{v}_{app}(\bm{x},t) = \bm{L}_{app}(t) \bm{x}(t).
     \label{eq: Applied Velocity}
\end{equation}
Deviation from the applied velocity field (see \textbf{Fig. \ref{SI: Velocity deviation map}} for sample heat maps) is then defined as:
\begin{equation}
    \delta \bm v = \bm v_{app}(\bm x,t) - \bm v(\bm x,t). 
    \label{eq: Velocity deviation}
\end{equation}

We additionally decompose the velocity gradient into symmetric and skew-symmetric components to characterize local raft expansion/compression and spin (in both directions) through tr$(\bm{D})$, and $\omega_{12}$, respectively. See \textbf{Figs. \ref{SI: Divergence heat map}} and \textbf{\ref{SI: Curl heat map}} for sample heat maps of local divergence and spin. Here, $\bm D$ is the rate of deformation tensor, $\bm{D}$, defined as:
\begin{equation}
    \bm{D} = \frac{1}{2} \bigl( \bm{L} + \bm{L}^T \bigr),
    \label{eq: Rate of Deformation}
\end{equation}
while $\bm \omega$ is the spin tensor:
\begin{equation}
    \bm{\omega} = \frac{1}{2} \bigl( \bm{L} - \bm{L}^T \bigr).
    \label{eq: Rate of Spin}
\end{equation}
Note that tr$(\bm{D})$ defines the true volumetric strain rate while $\bm{\omega}$ defines the network vorticity rate.

\subsection{Damage-based image analysis}
\label{sec: Methods: Damage-based Image Analysis}

Image anaysis was conducted on rafts loaded uniaxially to failure (\textbf{Fig. \ref{fig: damage-analysis}.A} and \textbf{\ref{SI: Uniaxial image analysis}}), as well as cyclically (\textbf{Fig. \ref{SI: Cyclic image analysis}}), to characterize their damage signatures (e.g., void onset and nucleation). First, colored images stacks were converted to binary using a manually adjusted color threshold in ImageJ \cite{schneider_nih_2012}, such that white pixels depicted ants, while black pixels depicted the apparatus, surrounding water, and the vacant spaces between ants. Once binary image stacks were obtained, further analysis was conducted using MATLAB R2022b. First, the four corners of the ant rafts (\textbf{Fig. \ref{fig: damage-analysis}.A} - red dots) - defining the domains in which image analysis would be conducted - were manually identified at the initial frame of mechanical loading. To account for non-normal camera alignment and perspective, an initial linear mapping was first applied that enforced an orthonormal basis of the boundaries (see basis $\{ \bm e_1, \bm e_2 \}$ in \textbf{Fig. \ref{fig: damage-analysis}.A}). To then track the domain boundaries through time, the deformation gradient, $\bm F(t) = $ diag$(1,\epsilon(t))$, applied during mechanical loading was used to update the domains' four corners in each subsequent frame (\textbf{Fig. \ref{fig: damage-analysis}.A} - red dotted lines). 

Once bounded, the quantities, positions (i.e., centroids denoted with cyan asterisks in \textbf{Fig. \ref{fig: damage-analysis}.A}), and areal distributions of voids within the domains were measured in time. Voids are here defined as defects in the raft that are too large for a fire to reach across or traverse, and which therefore cannot be immediately mended by either surface ants depositing into them, or raft ants binding with each other from opposite sides. Thus, two image processing steps were conducted to filter out voids that did not meet these size criteria. First, white pixels were dilated by a linear distance of $\ell/2$ such that any void with a gap dimension less than the body length of a single ant would be closed. This ensured that any gaps and fissures across which ants could walk or reach were filtered out. The white pixels were then eroded by the same length to restore the remaining voids to approximately their initial areas. Second, any voids whose areas, $A_v$, were less than the approximate reach, $A_{r} \approx \pi \ell^2$, enveloped by a single ant of length, $\ell$, were removed. This ensured that any voids that could be mended or closed by the deposition or positioning of a single ant, were also filtered out.   

The numbers, $N_v$, and average areas $A_v$ of voids were measured and normalized by the number of ants ($\sim 10^4$) and void area, $A_r$, respectively. Each of these values was then ensemble-averaged over $n=4$ experiments. The characteristic lag strain and time ($\epsilon_c$ and $\tau$) between when the peak number and area of voids occurred was then computed using the cross-correlational function:
\begin{equation}
    G(N_v^*,A_v^*) = \langle N_v^* (\epsilon) \cdot A_v^* (\epsilon + \delta \epsilon) \rangle - \langle N_v^* (\epsilon) \rangle \langle A_v^* (\epsilon + \delta \epsilon) \rangle,  
    \label{eq: Cross-correlation function}
\end{equation}
where the operator $\langle \square \rangle$ denotes the ensemble average over data within the strain range $\epsilon \in \left[ \delta \epsilon, \epsilon_{max}  - \delta \epsilon \right]$ (given maximum strain, $\epsilon_{max}$), and $\delta \epsilon$ is the incremental strain over which correlation is computed (see \textbf{Fig. \ref{SI: Cross-correlation}.D} for examples of $G$ with respect to $\delta \epsilon$). 

In addition to measuring and characterizing the void statistics, this binary image analysis was also used to coarsely estimate the minimum cross-sectional lengths, $L_{x}$ of continuous ants spanning direction $\bm e_1$ (e.g., solid red lines in \textbf{Figs. \ref{fig: damage-analysis}.A}, \textbf{\ref{SI: Uniaxial image analysis}}, and \textbf{\ref{SI: Cyclic image analysis}}). To do so, the raft domains were partitioned into a discrete set of pixel columns parallel to direction $\bm e_1$ and spaced approximately every $\ell$ (e.g., dotted red lines in \textbf{Fig. \textbf{\ref{SI: Cyclic image analysis} analysis}}). The column hosting the minimum number, $N_w$, of white pixels was then used to loosely estimate the instantaneous minimum cross-sectional length as $L_{x} \approx N_w dx/dp$, where $dp/dx$ is the unit conversion of pixels per unit length.


\bibliographystyle{rsc}

\newpage
\setcounter{figure}{0}    
\renewcommand{\thefigure}{s\arabic{figure}}

\setcounter{section}{0}    
\renewcommand{\thesection}{S\arabic{section}.}

\setcounter{equation}{0}    
\renewcommand{\theequation}{S\arabic{equation}}

\begin{center}
{\large \bf Electronic Supplementary Information for: \\
\textit{The mechanical response of fire ant rafts}}
\end{center}

\begin{center}
 {\bf Robert J. Wagner$^1$, Samuel Lamont$^2$, Zachary White$^2$, \& Franck J. Vernerey$^2$}
\end{center}

\begin{center}
$^1$Sibley School of Mechanical \& Aerosspace Engineering, \\
Cornell University, Ithaca, NY, USA\\
$^2$Paul M. Rady School of Mechanical Engineering, \\
University of Colorado, Boulder, CO, USA \\
\end{center}

\beginsupplement

\section{Consideration of drag}
\label{sec: SI: Drag}

One potential source of uncertainty on the force measurement is the viscous drag on Substrate 2 (see \textbf{Fig. \ref{SI: Exp setup}}), which would resist the tension applied to said substrate by the ants, and therefore diminish the elongation of the calibrated elastic spring. The magnitude of drag force on the block may be estimated as:

\begin{equation}
    F_d = \frac{1}{2}\rho v^2 c_d A,
    \label{eq:Drag Force}
\end{equation}

where $\rho = 1\times10^3$ [kg m$^{-3}$] is the density of water, $v \approx 1 \times 10^{-3}$ [mm s$^{-1}$] is the approximate peak velocity, $c_d \approx 0.7$ is the approximate drag coefficient of a rectangular body, and $A \approx 4 \times 10^{-3}$ [m$^3$] is said substrates approximate cross-sectional area. Substituting these values into Eqn. (\ref{eq:Drag Force}) we find that the order of drag force is 0.1 dynes, Which is four to five orders of magnitude smaller than the net reaction force measured in the ant raft. Indeed, even if we unrealistically supposed that Substrate 2 traveled at the peak pulling rate of $1\times 10^{-2}$ [m s$^{-1}$], then drag force would still remain on the order of 10 dynes, or three orders of magnitude smaller than the measured forces transmitted by the ant network. 

While the drag force on the ant raft itself was also considered, it is considerably more difficult to estimate. However, because of the loading configuration, drag on the raft would slow transmission of force to Substrate 2 and cause only a reduction of measured force with respect to increasing strain rate. Yet measured spring forces and estimated raft stiffness are consistently greater as strain rate is increased, thus suggesting that the key finding pertaining to stress response - that fire ant rafts exhibit rate-dependent stiffening behavior - could only be exacerbated by in the absence of drag forces. For these reasons, we conclude that drag forces do not obfuscate the key findings of this paper. Notably, at the highest two strain rates, softening - attributed to premature raft failure - was observed (see \textbf{Fig. \ref{SI: Delamination}}). Drag forces are one potential source of this failure, and so this data is not presented in the main manuscript.

\figuremacro{H}{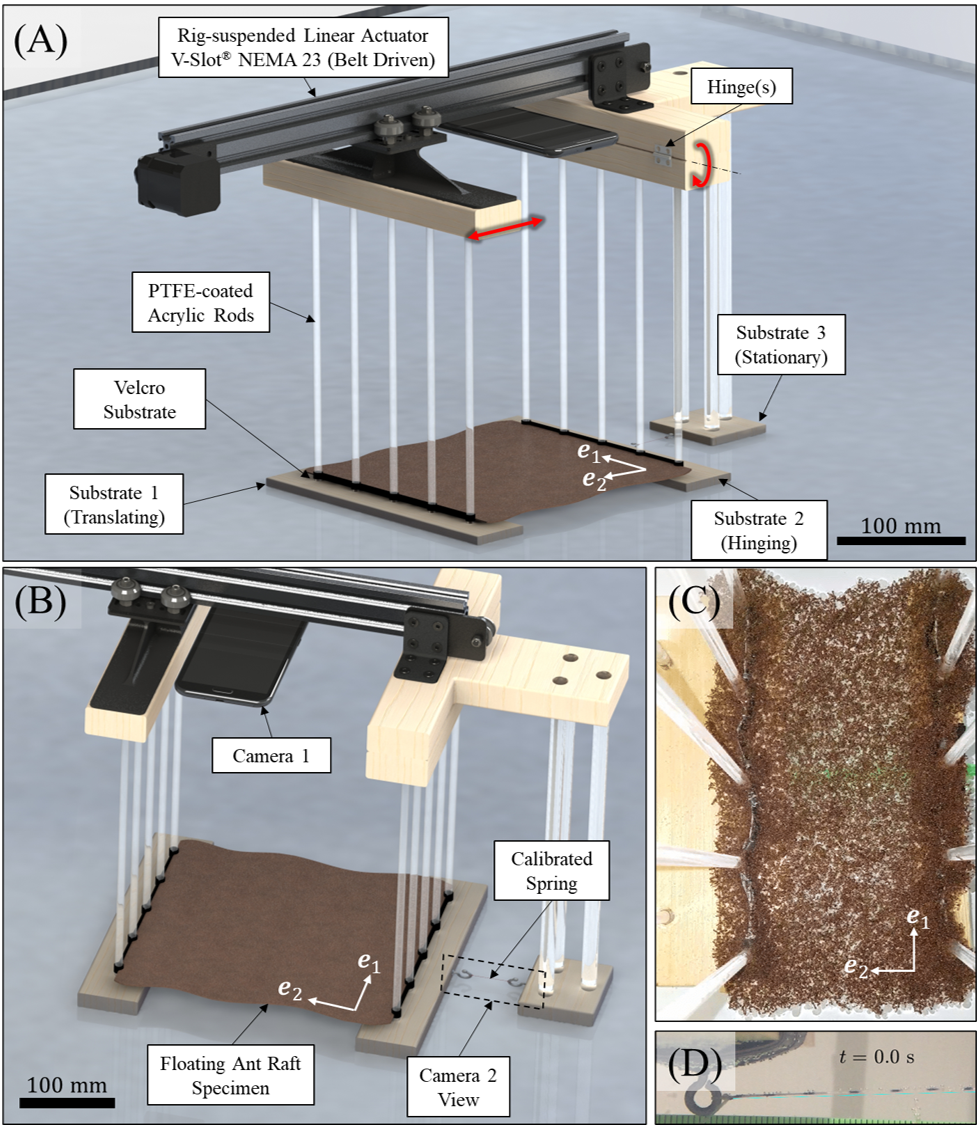}{}{\textbf{Experimental apparatus.} \textbf{(A-B)} Two isometric views of a 3D model of the tensile testing apparatus are shown. \textbf{(A)} Denotes the basic mechanical components of the apparatus while \textbf{(B)} denotes the two camera positions or views used to capture footage, as well as the ant raft specimen position, and the location of the calibrated elastomeric spring used to estimate raft reaction force. The two-headed red arrow denotes the translation of the linear actuator platform, to which Substrate 1 is rigidly attached. The circular red arrow shows the axis about which Substrate 2 hinges, as well as the direction it hinges due to experimental loading. \textbf{(C-D)} Sample photographs obtained from \textbf{(C)} Camera 1 and \textbf{(D)} Camera 2 are shown. White arrows denote the orthonormal basis, $\{\bm{e}_1,\bm{e}_2 \}$ where $\bm{e}_2$ is the principal loading direction. \label{SI: Exp setup}}{0.7}

\figuremacro{H}{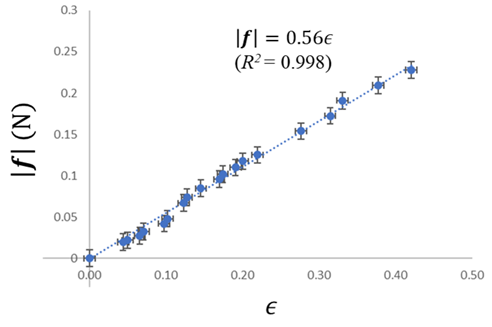}{}{\textbf{Elastic band calibration.} Calibrated force-strain relation of the wet elastic spring used to estimate transmitted load by the ants. The spring is roughly linear ($R^2 = 0.998$) and has a spring constant of 0.56 N per unit strain. \label{SI: Force calibration}}{0.4}

\figuremacro{H}{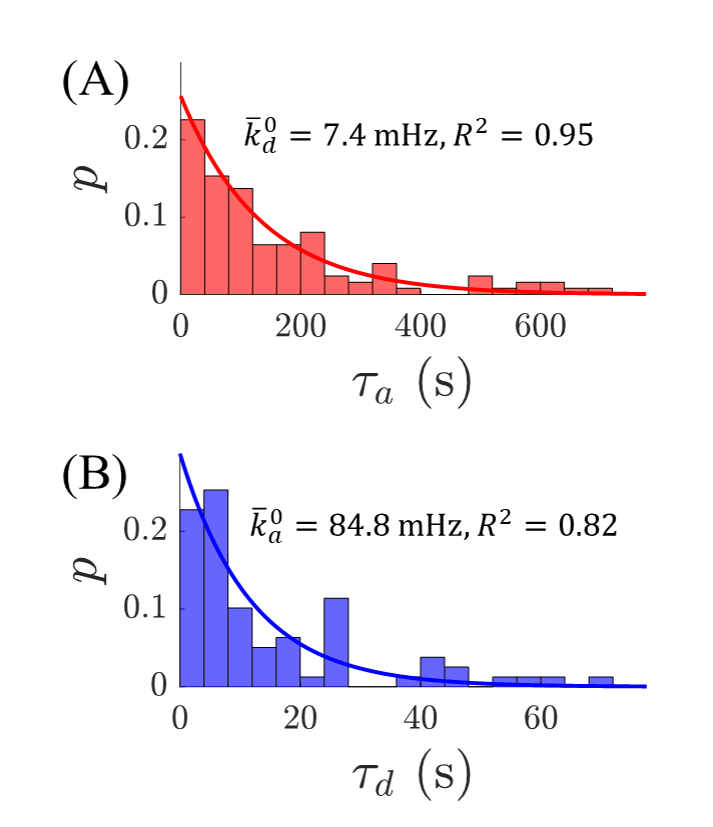}{}{\textbf{Open and closed bond lifetime distributions.} \textbf{(A)} The probability distribution function of an ant-to-ant bond’s lifetime is plotted with the red curve representing the exponential decay function of the form $P_d =\exp (-\bar k_d^0 \tau_a)$. The average detachment rate in an unperturbed raft is estimated as $\bar k_d^0 = 7.4$ mHz ($R^2 =0.95 $). \textbf{(B)} The probability distribution function of an ant leg’s detached lifetime is plotted with the blue curve representing the exponential decay function of the form $P_a = \exp (-\bar k_a^0 \tau_d)$. The average attachment rate in an unperturbed raft is estimated as $\bar k_a^0 = 84.8$ mHz ($R^2 =0.82 $). \label{SI: Bond dynamics}}{0.45}

\figuremacro{H}{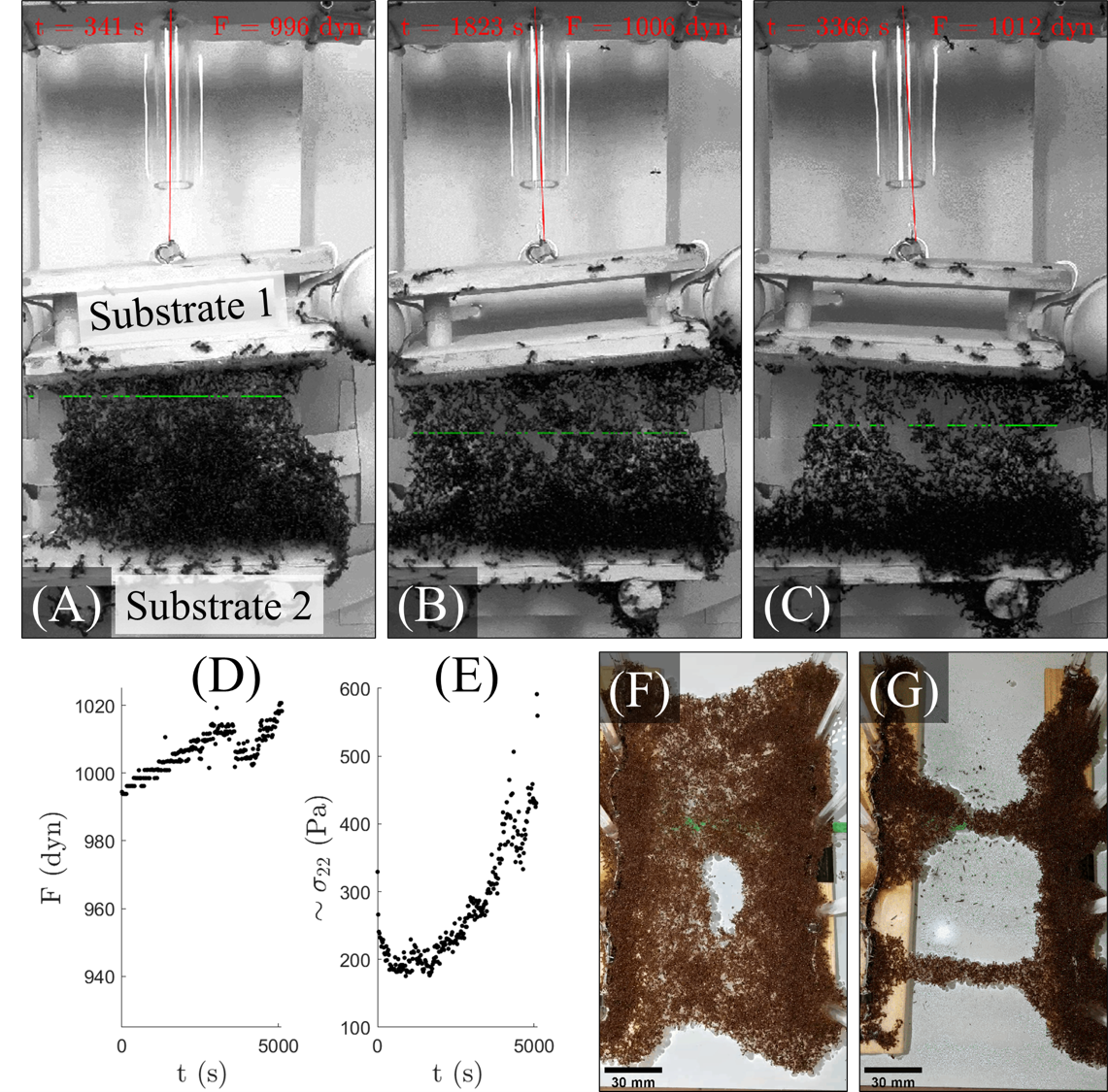}{}{\textbf{Active contraction drives dissolution.} \textbf{(A-C)} An apparatus capable of loosely estimating active contractile force is depicted at frames roughly 1500 seconds apart, highlighting the steady onset of vacant raft regions due to ants' tendency to aggregate at dry zones on the apparatus rather than replenish the raft layer. The apparatus is comprised of a floating substrate (Substrate 1) anchored in place by the tension in an elastic band (red) of known force-extension relation (see \textbf{Fig. S2} for example). The ant raft is suspended between Substrate 1 and a second, fixed substrate (Substrate 2) so that any active contractile force generated by the ants displaces Substrate 1 and stretches the spring so that \textbf{(D)} contractile force may be estimated in time. The green line highlights the minimum cross-sectional raft length (multiplied by the characteristic thickness of an ant to obtain area) in the direction transverse to contractile stress allowing for very rough estimate of \textbf{(E)} raft stress. Panels \textbf{(F-G)} illustrate the same phenomena occurring in the apparatus used for mechanical testing over a duration of roughly 1 hour. \label{SI: Active contraction}}{0.65}

\figuremacro{H}{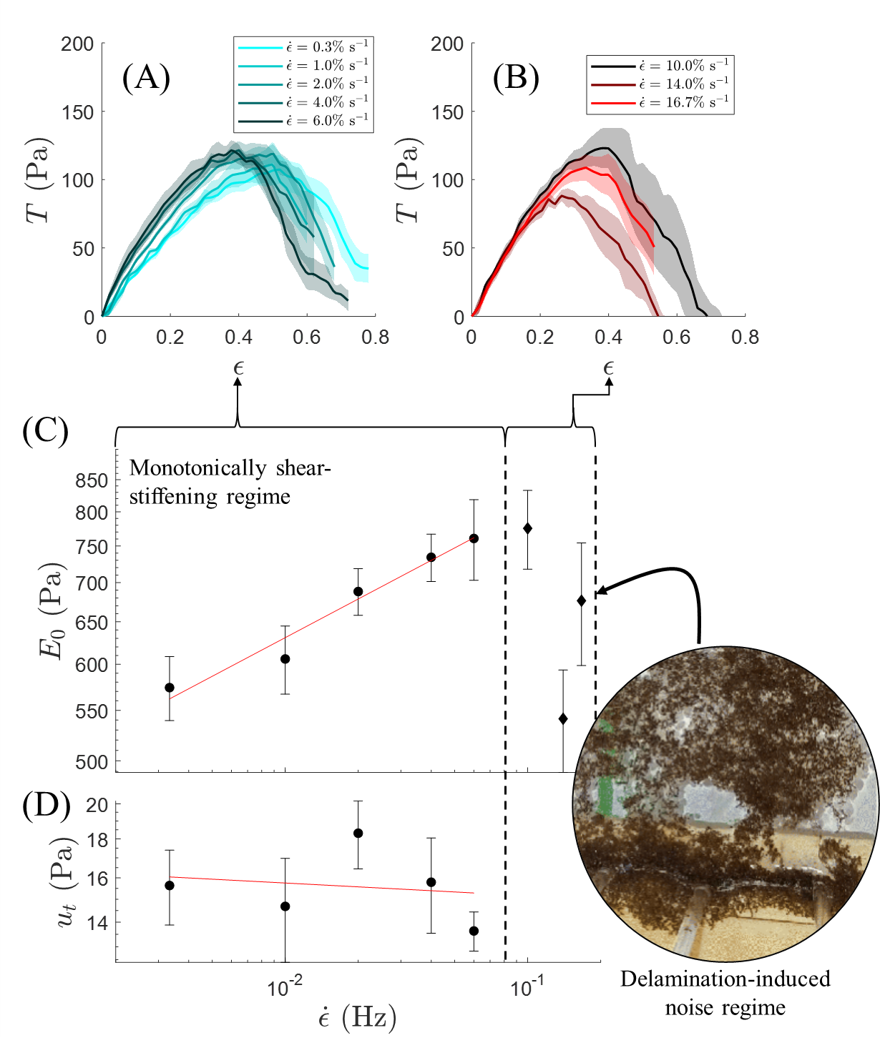}{}{\textbf{Drag-induced delamination at high strain rates.} \textbf{(A-B)} Ensemble-averaged ($n=4$) engineering stress versus strain for \textbf{(A)} $\dot \epsilon = \{0.3,1.0,2.0,4.0,6.0 \} \%$ s$^{-1}$ and \textbf{(B)} $\dot \epsilon = \{10.0,14.0,16.7 \} \%$ s$^{-1}$. Shaded regions represent S.E. \textbf{(C)} Plotting initial tangent modulus with respect to strain rate reveals a regime of monotonically increasing stiffness and a second regime for which there is no discernible correlation between strain rate and raft stiffness. Examination of the videos reveals that at strain rates of $\dot \epsilon \geq 10 \%$ s$^{-1}$, delamination at the ant-substrate interface occurred for many of the experiments. \textbf{(D)} Mechanical toughness for the rats loaded in the monotonically shear-stiffening regime is reported ($u_t \propto \dot \epsilon^{-0.02}$, $R^2 = 0.03$). \label{SI: Delamination}}{0.7}

\figuremacro{H}{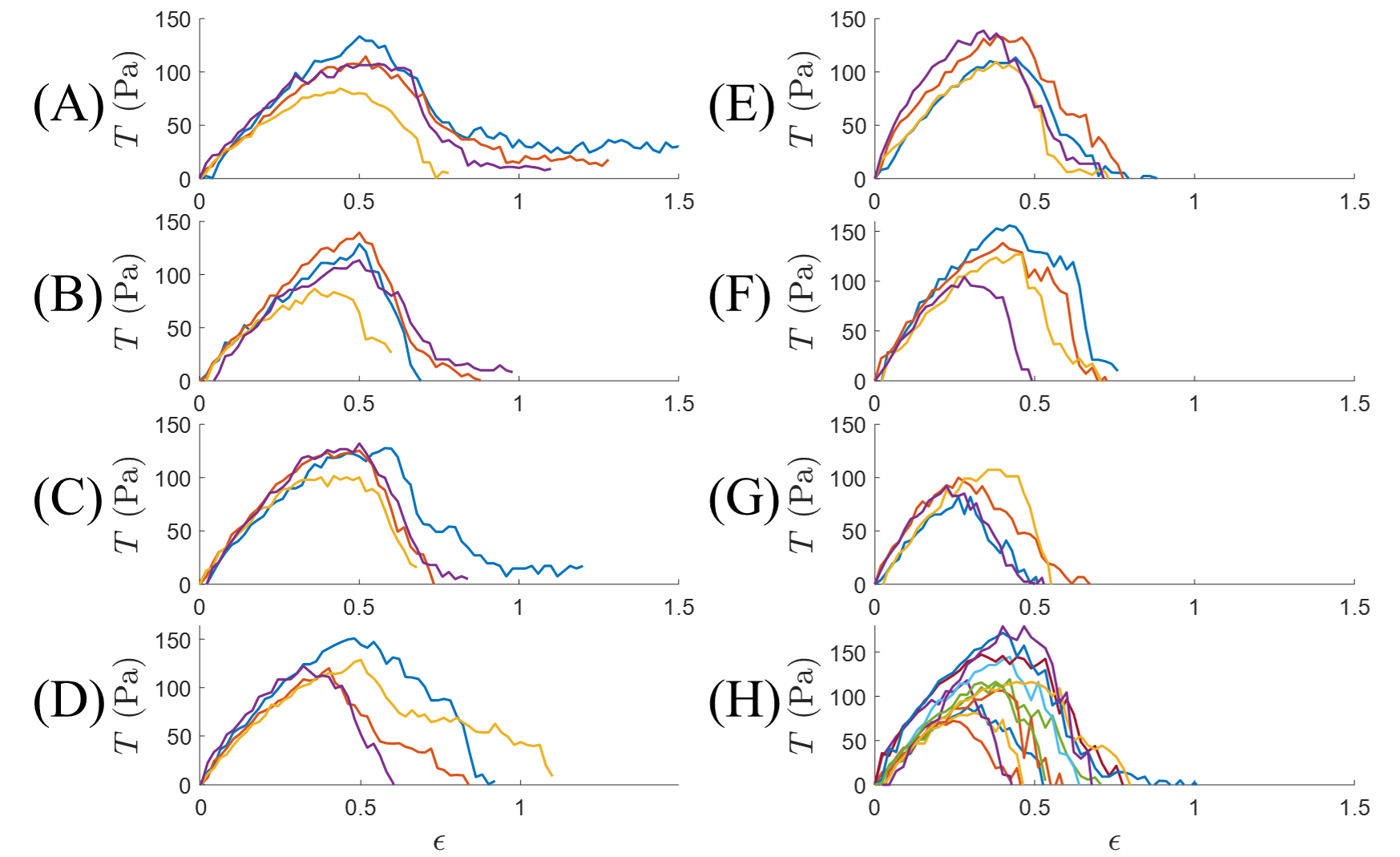}{}{\textbf{Extended stress-strain data.} Raw engineering stress-strain data is shown for applied strain rates of \textbf{(A)} $0.3 \%$ s$^{-1}$, \textbf{(B)} $1 \%$ s$^{-1}$, \textbf{(C)} $2 \%$ s$^{-1}$, \textbf{(D)} $4 \%$ s$^{-1}$, \textbf{(E)} $6 \%$ s$^{-1}$, \textbf{(F)} $10 \%$ s$^{-1}$, \textbf{(G)} $14 \%$ s$^{-1}$, and \textbf{(H)} $16.7 \%$ s$^{-1}$. \label{SI: Extended stress}}{0.7}

\figuremacro{H}{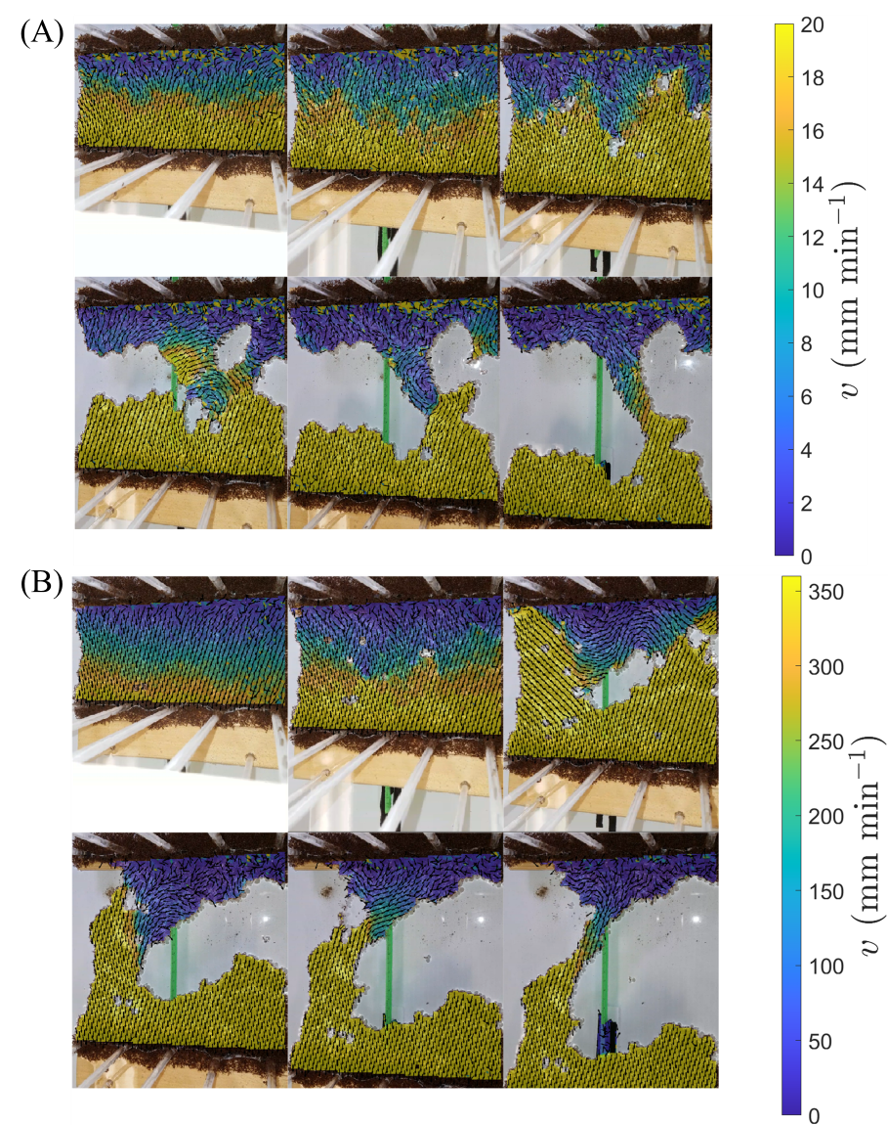}{}{\textbf{Examples of measured velocity.} Heat maps of the the measured raft speed, $v$, for applied strain rates of \textbf{(A)} $0.3 \%$ s$^{-1}$ (20 mm min$^{-1}$ extension speed) and \textbf{(B)} $6 \%$ s$^{-1}$ (360 mm min$^{-1}$ extension speed). Samples are initially $L_0 = 100$ mm or $\sim 34 \ell$ wide in the loading direction (vertical axis). The vector fields represent the normalized order parameter, $\hat{\bm \varphi}$. Panels are depicted from the moment of initial loading to just before ultimate failure in intervals of approximately $17\%$ of the total failure strain.  \label{SI: Velocity heat map}}{0.85}

\figuremacro{H}{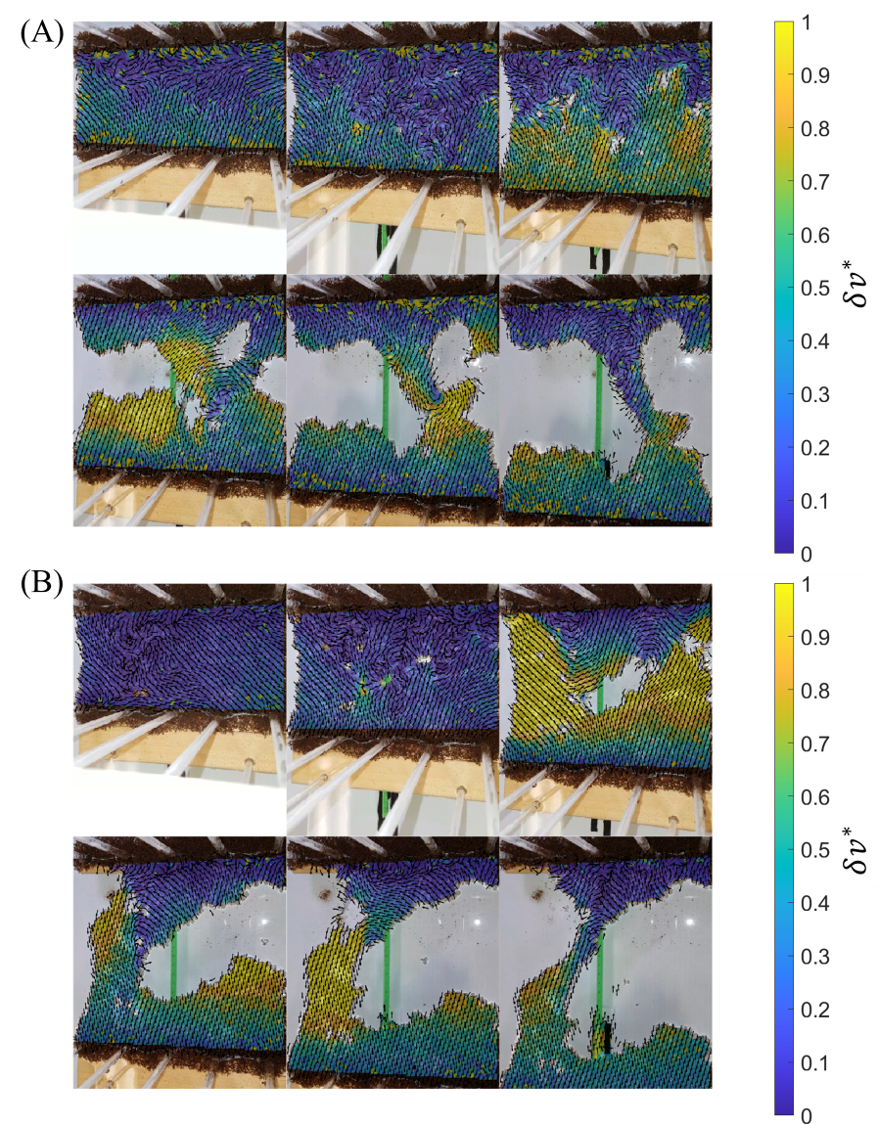}{}{\textbf{Examples of velocity deviation from applied conditions.} Heat maps of the normalized corrected (i.e., non-affine) speed, $\delta v^* = |\bm v - \bm v_{app}|/(\dot \epsilon L_0)$, at applied strain rates of \textbf{(A)} $0.3 \%$ s$^{-1}$ (20 mm min$^{-1}$ extension speed) and \textbf{(B)} $6 \%$ s$^{-1}$ (360 mm min$^{-1}$ extension speed). Regions where $\delta v^*\approx 1$ are deviating from the applied velocity at approximately the loading speed. Samples are initially $L_0 = 100$ mm or $\sim 34 \ell$ wide in the loading direction (vertical axis). The vector fields represent the normalized order parameter, $\hat{\bm \varphi}$. Panels are depicted from the moment of initial loading to just before ultimate failure in intervals of approximately $17\%$ of the total failure strain. \label{SI: Velocity deviation map}}{0.85}

\figuremacro{H}{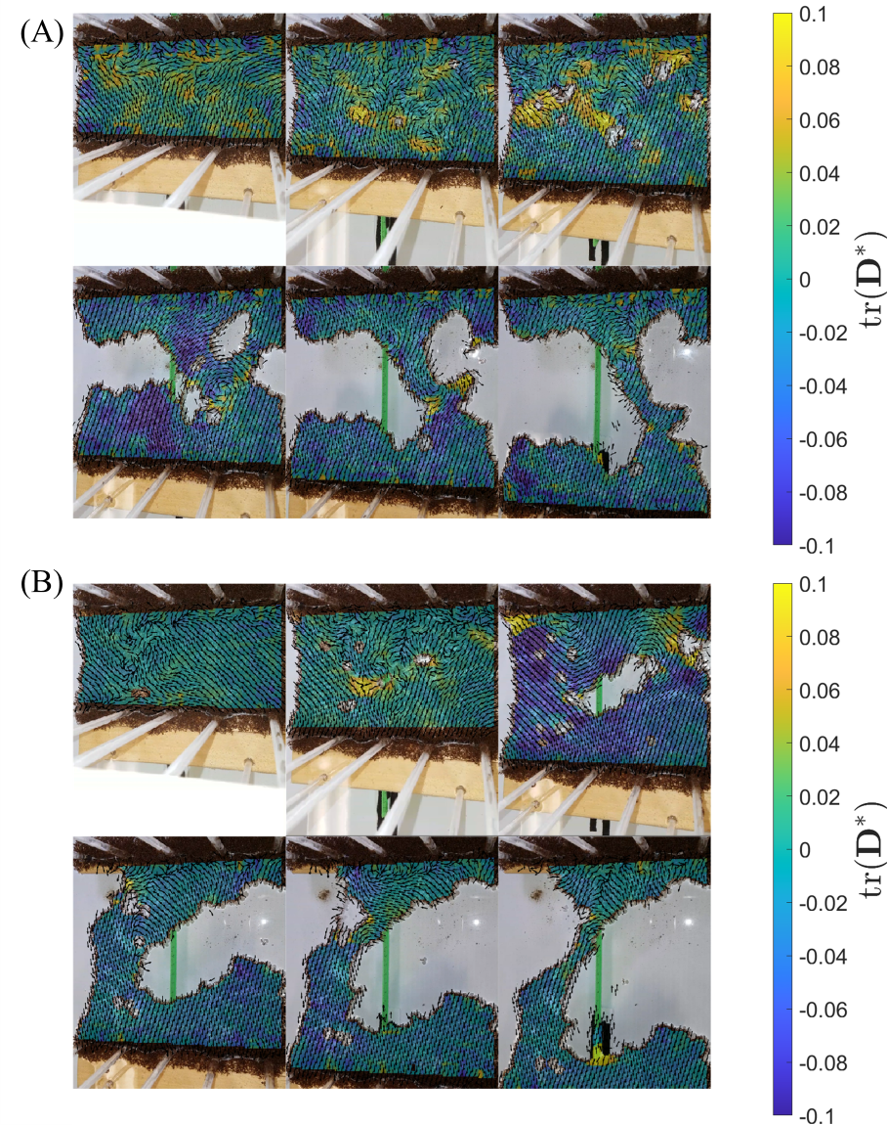}{}{\textbf{Examples of rate of expansion/contraction.} Heat maps of the normalized rate of expansion/contraction, tr$(\bm D^*)$, at applied strain rates of \textbf{(A)} $0.3 \%$ s$^{-1}$ (20 mm min$^{-1}$ extension speed) and \textbf{(B)} $6 \%$ s$^{-1}$ (360 mm min$^{-1}$ extension speed). tr$(\bm D^*)>0$ denotes regions of local expansion, while tr$(\bm D^*)<0$ denotes local contraction. Samples are initially $L_0 = 100$ mm or $\sim 34 \ell$ wide in the loading direction (vertical axis). The vector fields represent the normalized order parameter, $\hat{\bm \varphi}$. Panels are depicted from the moment of initial loading to just before ultimate failure in intervals of approximately $17\%$ of the total failure strain. \label{SI: Divergence heat map}}{0.85}

\figuremacro{H}{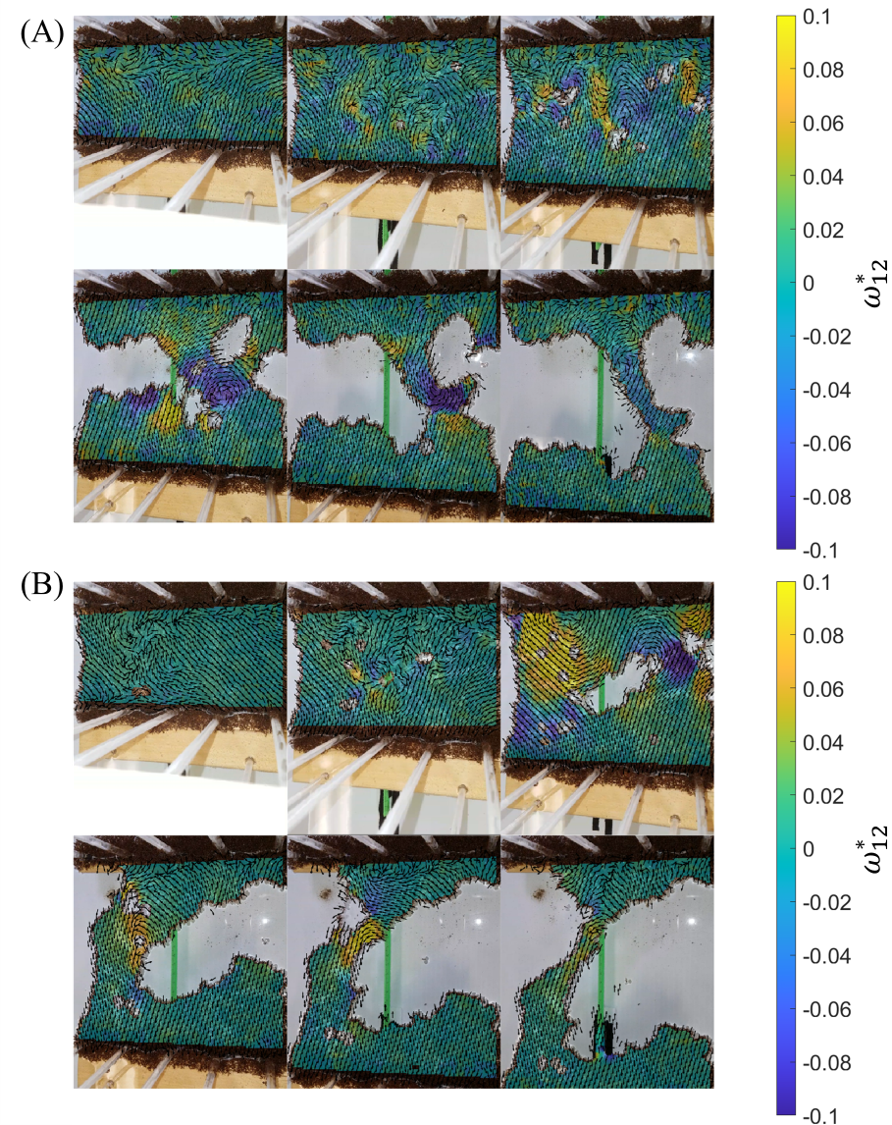}{}{\textbf{Examples of rate of spin.} Heat maps of the normalized rate of spin, $\omega_{12}^*$, at applied strain rates of \textbf{(A)} $0.3 \%$ s$^{-1}$ (20 mm min$^{-1}$ extension speed) and \textbf{(B)} $6 \%$ s$^{-1}$ (360 mm min$^{-1}$ extension speed). $\omega_{12}^*<0$ denotes regions of local clockwise spin, while $\omega_{12}^*>0$ denotes regions of local counter-clockwise spin. Samples are initially $L_0 = 100$ mm or $\sim 34 \ell$ wide in the loading direction (vertical axis). The vector fields represent the normalized order parameter, $\hat{\bm \varphi}$. Panels are depicted from the moment of initial loading to just before ultimate failure in intervals of approximately $17\%$ of the total failure strain.  \label{SI: Curl heat map}}{0.85}

\figuremacro{H}{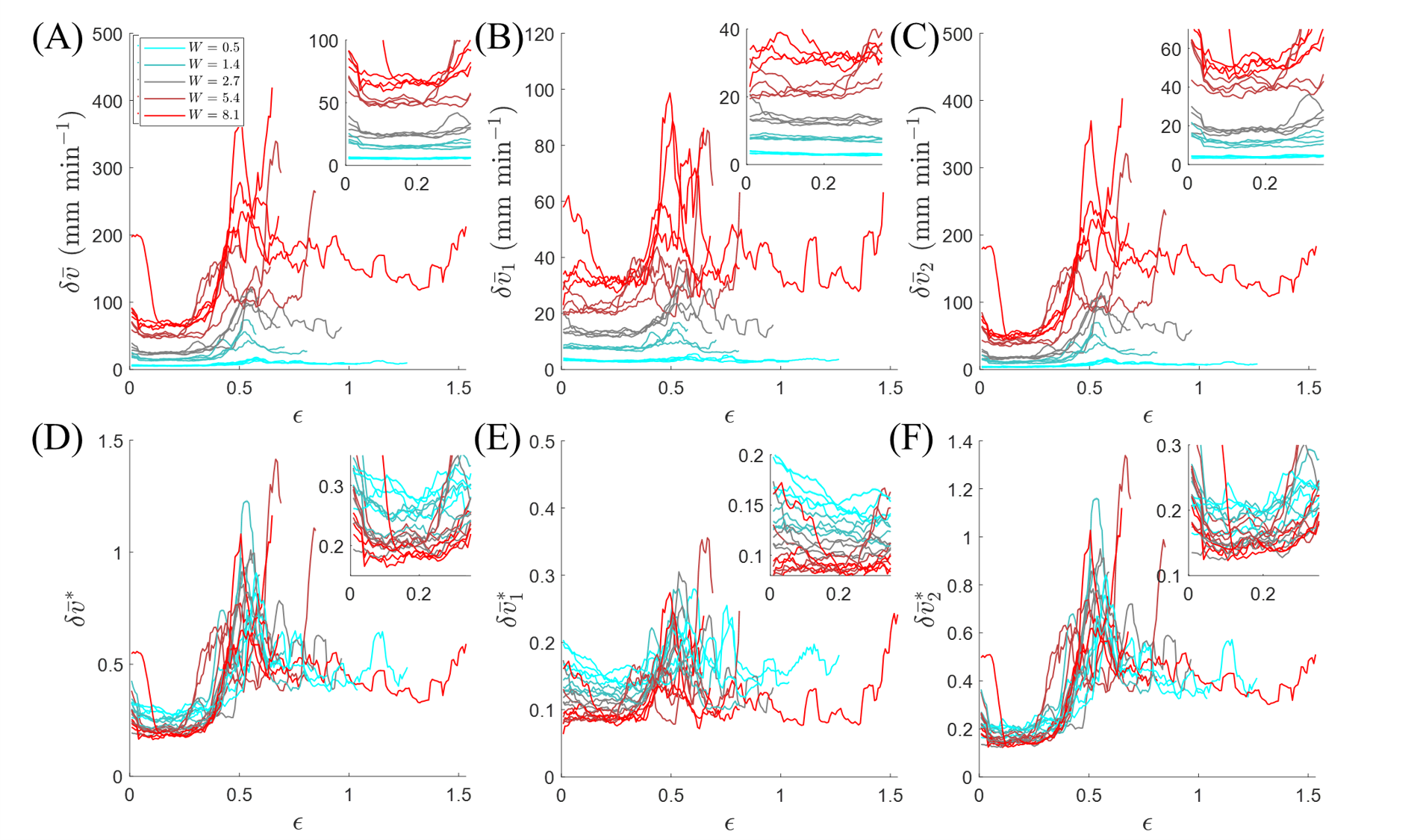}{Extended raw data for velocity deviation measures. }{\textbf{(A-C)} Deviation, $\delta \bar v$, and \textbf{(D-F)} normalized deviation, $\delta \bar v^*=\delta \bar v/(\dot \epsilon L_0)$, from the applied speed are plotted with respect to engineering strain for $n=4$ samples each at Weissenberg numbers of $W=\{0.5, 1.4, 2.7, 5.4, 8.1\}$. \textbf{(A,D)} Net values, and components in the directions \textbf{(B,E)} normal to and \textbf{(C,F)} in-line with the applied loading are presented. \label{SI: Nonaffine speed}}{1}

\figuremacro{H}{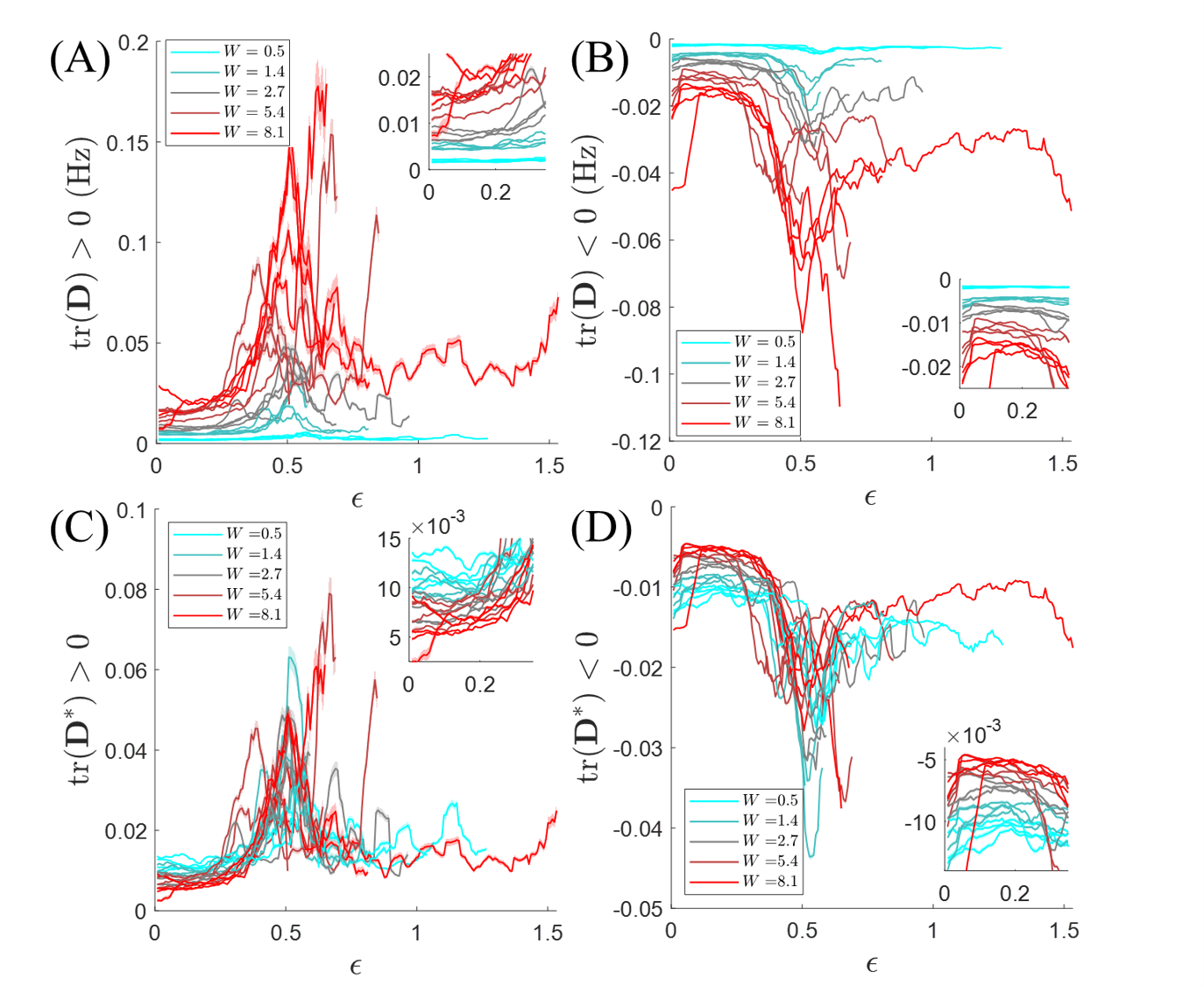}{}{\textbf{Extended raw data for rate of expansion/contraction.} \textbf{(A-B)} Spatiotemporally averaged \textbf{(A)} expansion, tr$(\bm D) >0$, and \textbf{(B)} contraction, tr$(\bm D)<0$, are plotted with respect to engineering strain for $n=4$ samples each at Weissenberg numbers of $W=\{0.5, 1.4, 2.7, 5.4, 8.1\}$. \textbf{(C-D)} The same plots are provided for the normalized rate of expansion/contraction, $\bm D^* = \bm D/\dot \epsilon$. Insets depict the same data of each respective plot for $\epsilon<0.35$, which is typically the regime in which no major damage (yet rate-dependent mechanical response) was observed. \label{SI: Raw expansion-contraction}}{0.75}

\figuremacro{H}{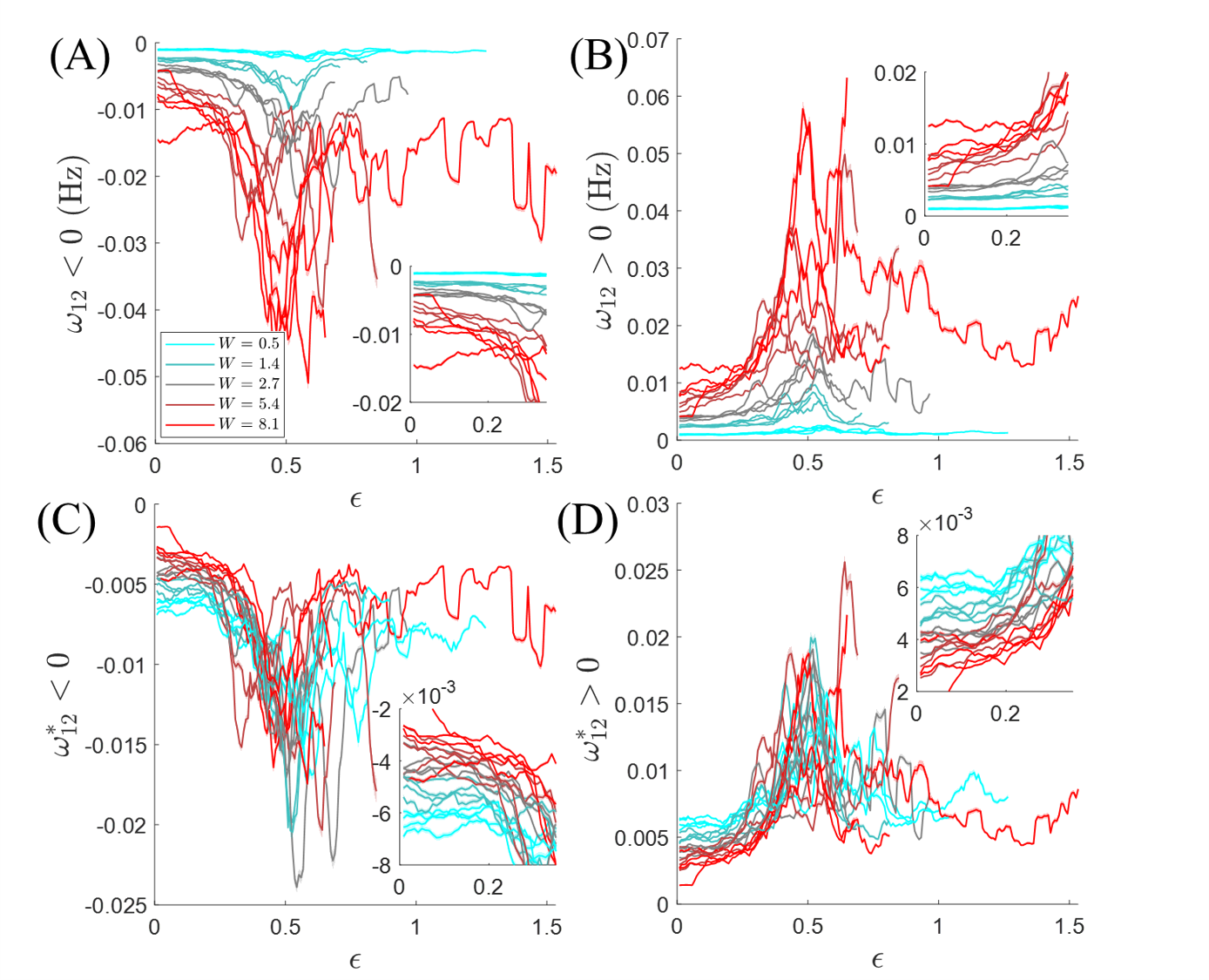}{}{\textbf{Extended raw data for rate of spin.} \textbf{(A-B)} Spatiotemporally averaged \textbf{(A)} clockwise, $\omega_{12} < 0$, and \textbf{(B)} counter-clockwise, $\omega_{12} > 0$, rates of spin are plotted with respect to engineering strain for $n=4$ samples each at Weissenberg numbers of $W=\{0.5, 1.4, 2.7, 5.4, 8.1\}$. \textbf{(C-D)} The same plots are provided for the normalized rates of spin, $\bm \omega^* = \bm \omega/\dot \epsilon$. Insets depict the same data of each respective plot for $\epsilon<0.35$, which is typically the regime in which no major damage (yet rate-dependent mechanical response) was observed. \label{SI: Raw spin}}{0.75}

\figuremacro{H}{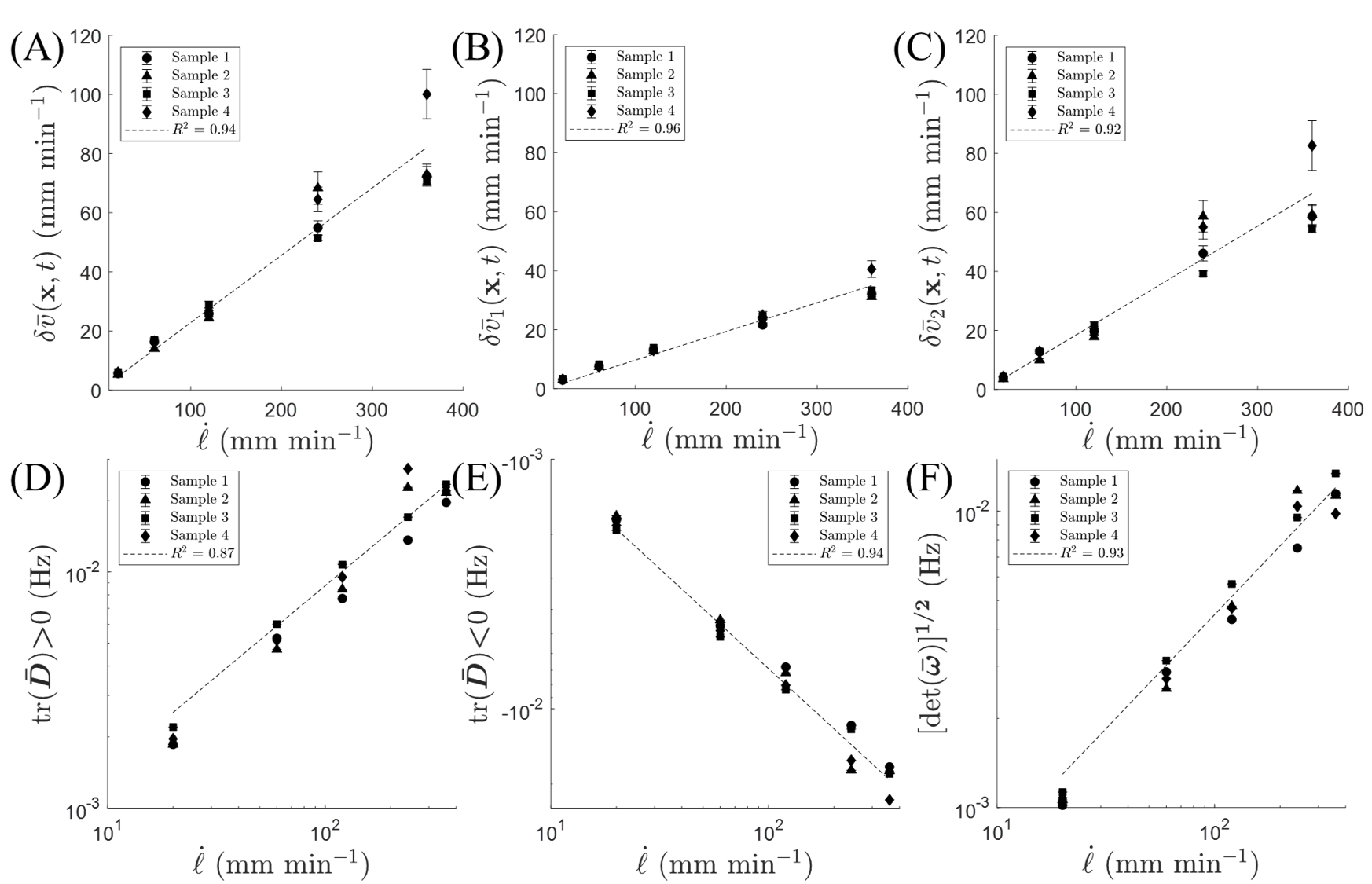}{}{\textbf{Non-normalized measures of rate-dependent strain localization.} \textbf{(A-C)} Spatiotemporally averaged deviations of speed from the applied conditions (for $\epsilon \leq 0.3$) with respect to applied loading speed. \textbf{(A)} Net speed, as well as speed in the directions \textbf{(B)} in-line with and \textbf{(C)} normal to the loading direction are shown separately. \textbf{(D-E)} Spatiotemporally averaged rates of local \textbf{(D)} expansion, \textbf{(E)} contraction, and \textbf{(F)} spin (in either direction) with respect to loading speed. \label{SI: Non-normalized localization}}{1}

\figuremacro{H}{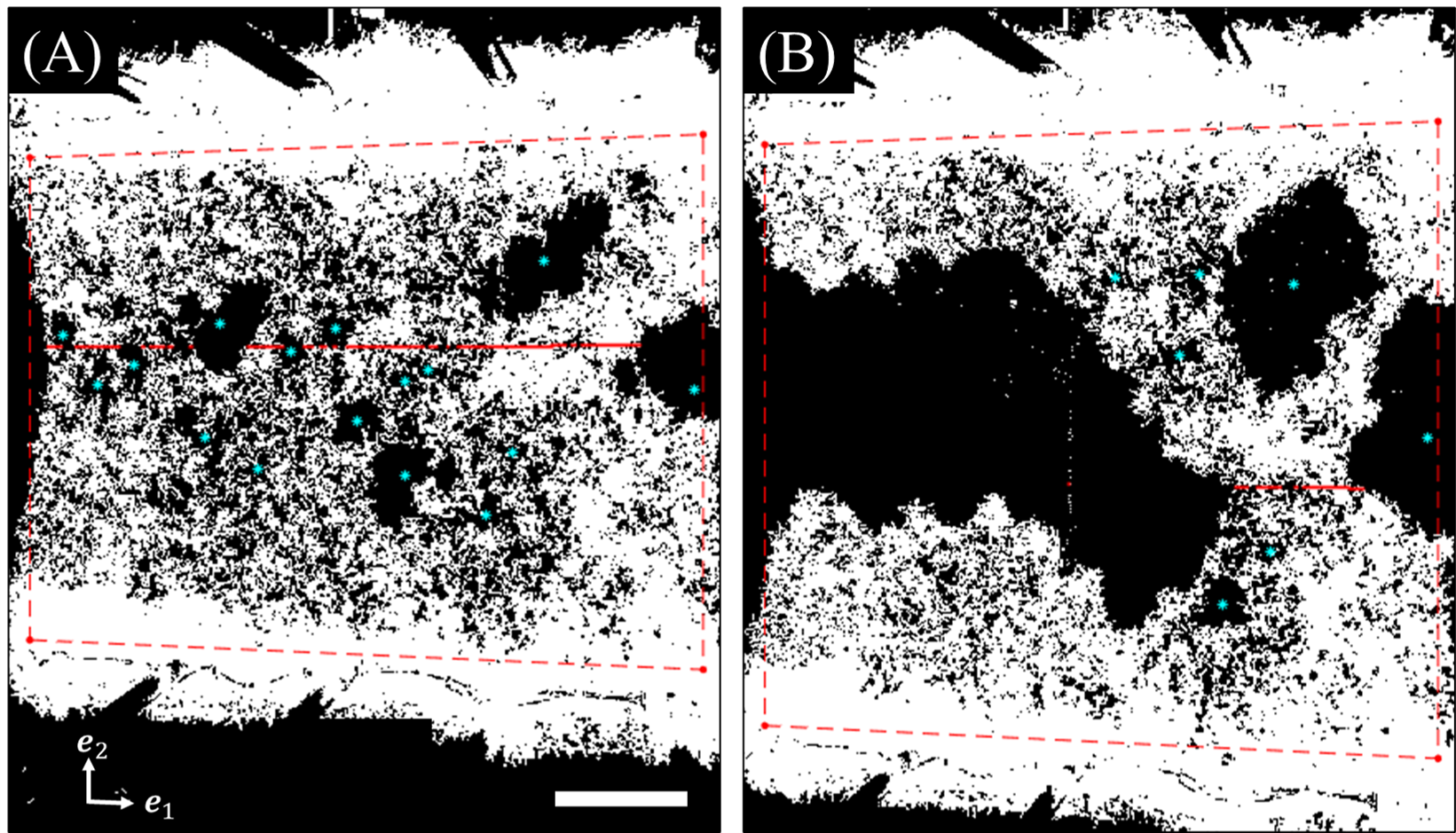}{}{\textbf{Annotated image analysis of uniaxial tension to failure.} \textbf{(A-B)} Binary image analysis of an ant raft in which the ants are depicted white while voids in the raft are depicted black \textbf{(A)} before and \textbf{(B)} during uniaxial loading to failure. The four red points connected by red dotted lines demark the corners of the orthonormal region of interest, which elongates at the prescribed extension rate, $\dot{\epsilon}$; and is used to interpolate the end positions of cross sections parallel to $\bm{e}_1$ (i.e., normally to the principal loading direction, $\bm{e}_2$), which adjusts for the camera’s non-normal perspective. Cyan points denote the centers of void spaces (i.e., continuous regions of black pixels). Scale bar in \textbf{(A)} represents $10 \ell$. \label{SI: Uniaxial image analysis}}{0.7}

\figuremacro{H}{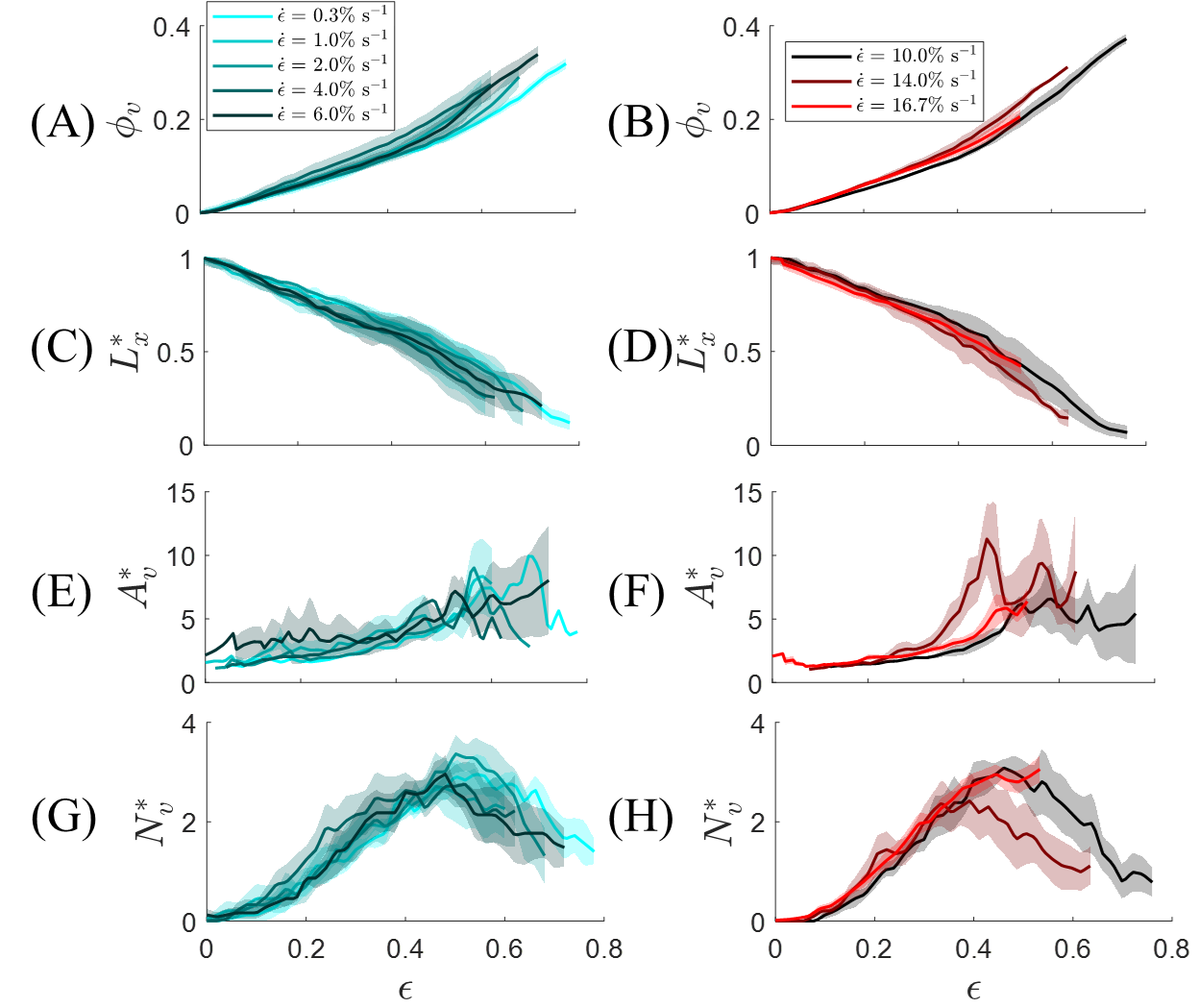}{}{\textbf{Extended ensemble-averaged damage analysis results.} \textbf{(A-B)} Raw change in areal free volume, $\phi_v = \phi-\phi_0$, \textbf{(C-D)} minimum cross sectional length, $L_x^*$, \textbf{(E-F)} average void area, $A_v^*$ (in units of $\pi \ell^2$), and \textbf{(G-H)} number of voids, $N_v^*$ (in units of $10^{-3}$ voids per ant) are provided for the strain rates both \textbf{(A,C,E,G)} without and \textbf{(B,D,F,H)} with excessive delamination. Shaded regions represent S.E. \label{SI: Extended ensemble average damage}}{0.7}

\figuremacro{H}{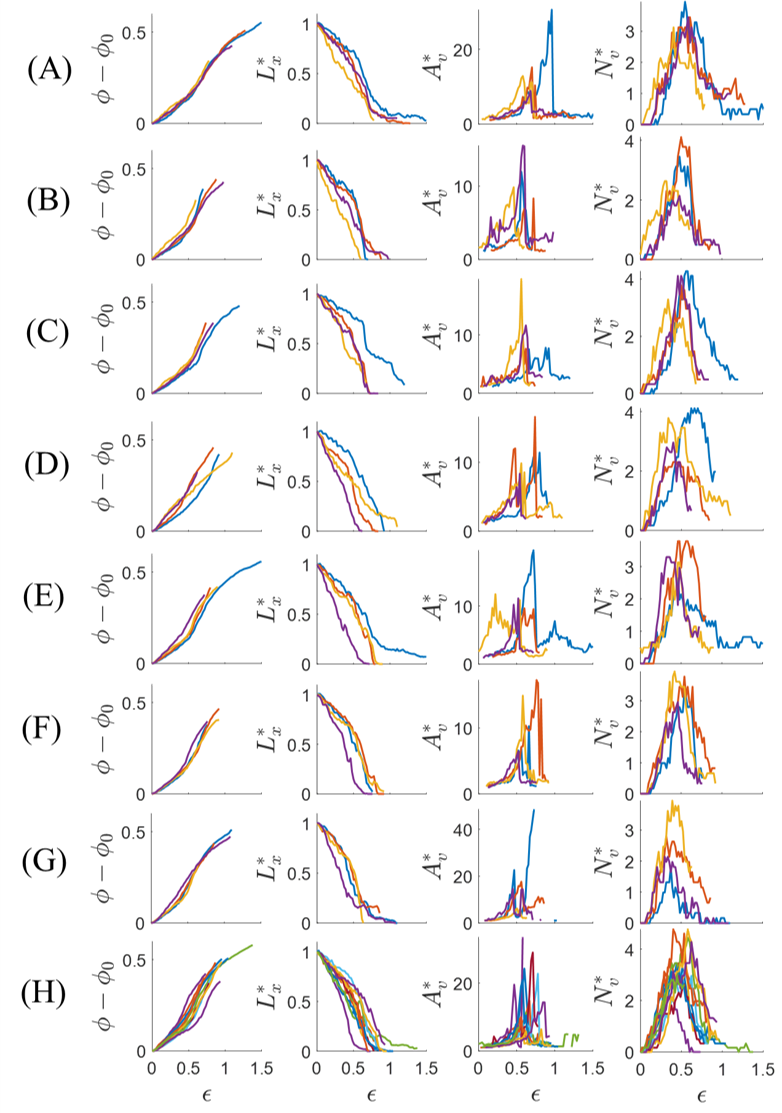}{}{\textbf{Extended image analysis results.} Raw change in areal free volume, $\phi_v = \phi-\phi_0$, minimum cross sectional length, $L_x^*$, average void area, $A_v^*$ (in units of $\pi \ell^2$), and number of voids, $N_v^*$ (in units of $10^{-3}$ voids per ant) are provided for \textbf{(A)} $0.3 \%$ s$^{-1}$, \textbf{(B)} $1 \%$ s$^{-1}$, \textbf{(C)} $2 \%$ s$^{-1}$, \textbf{(D)} $4 \%$ s$^{-1}$, \textbf{(E)} $6 \%$ s$^{-1}$, \textbf{(F)} $10 \%$ s$^{-1}$, \textbf{(G)} $14 \%$ s$^{-1}$, and \textbf{(H)} $16.7 \%$ s$^{-1}$. \label{SI: Extended raw damage}}{0.675}

\figuremacro{H}{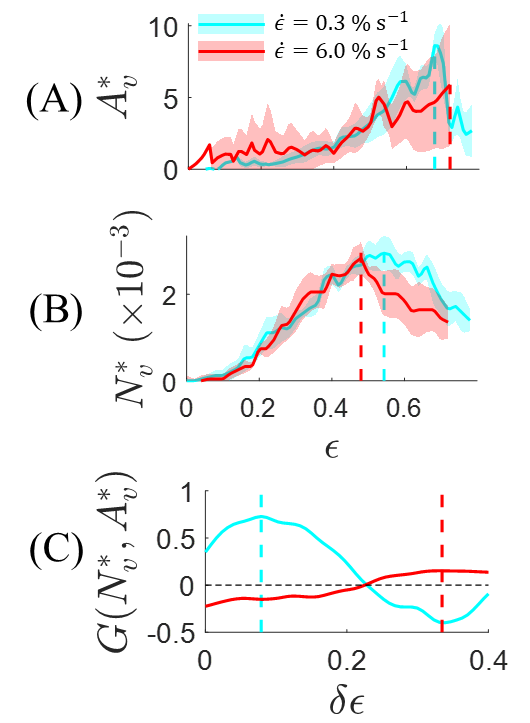}{Cross-correlation between $N_v^*$ and $A_v^*$. }{\textbf{(A)} average void area, $A_v^*$ (in units of $\pi \ell^2$), and \textbf{(B)} number of voids, $N_v^*$ (in units of $10^{-3}$ voids per ant) are provided for $\dot \epsilon = \{0.2,6 \} \%$ s$^{-1}$. \textbf{(C)} The corresponding cross-correlation function, $G$, is plotted with respect to incremental strain, $\delta \epsilon$, and the peaks of $G$ are used to determine the coalescence lag strains, $\epsilon_c$, for each loading rate (demarked by vertical dotted lines). Shaded regions represent S.E. \label{SI: Cross-correlation}}{0.4}

\figuremacro{H}{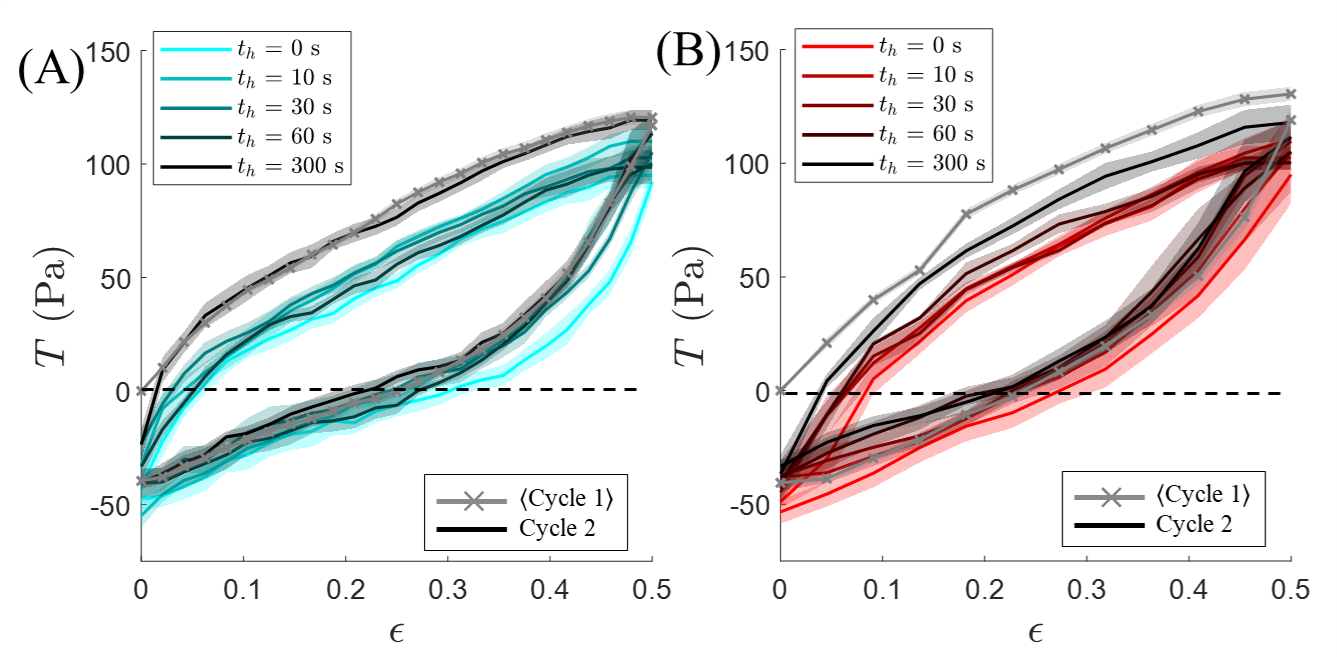}{Extended ensemble-averaged stress response for cyclic loading. }{\textbf{(A-B)} ensemble-averaged ($n=4$) engineering stress versus strain (to $50\%$) over two loading cycles with $t_h \in \{ 0, 10, 30, 60, 300 \}$ seconds for \textbf{(A)} $W=0.45$ and \textbf{(B)} $W=5.4$. Shaded regions represent S.E. \label{SI: ensemble-averaged cyclic stress}}{0.8}

\figuremacro{H}{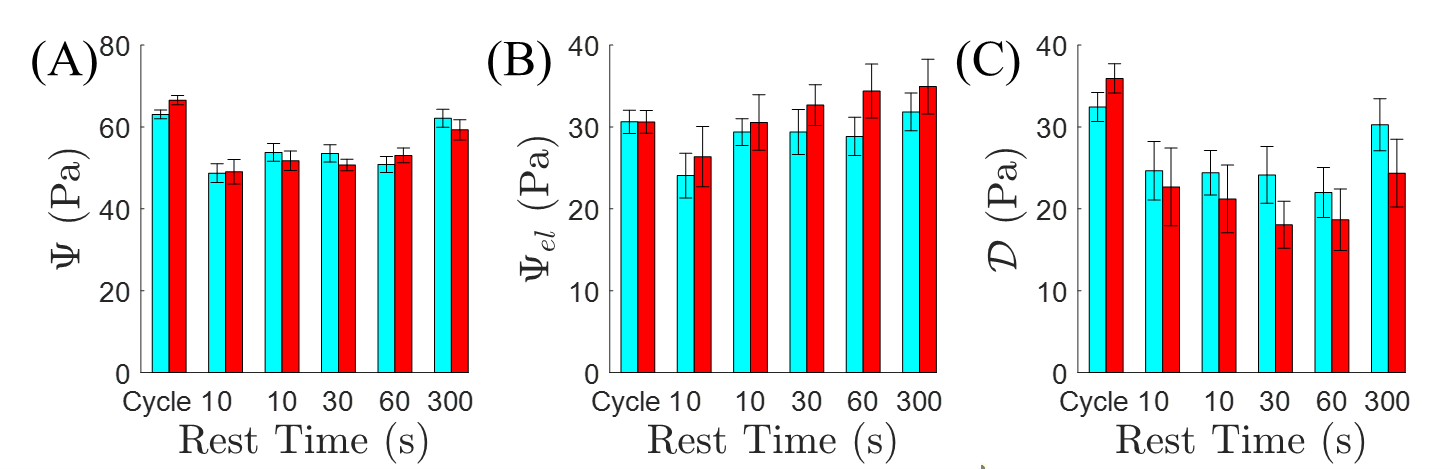}{Extended ensemble-averaged strain, stored, and dissipated energies. }{\textbf{(A-C)} Ensemble-averaged \textbf{(A)} strain energy, $\Psi$, \textbf{(B)} stored energy, $\Psi_{el}$, and \textbf{(C)} dissipated energy, $\mathcal{D}$, for recovery times of $t_h={0,300}$ s when $W=0.45$ (cyan bars) and $W=5.4$ (red bars). Error bars represent S.E. \label{SI: Extended strain energy}}{0.85}

\figuremacro{H}{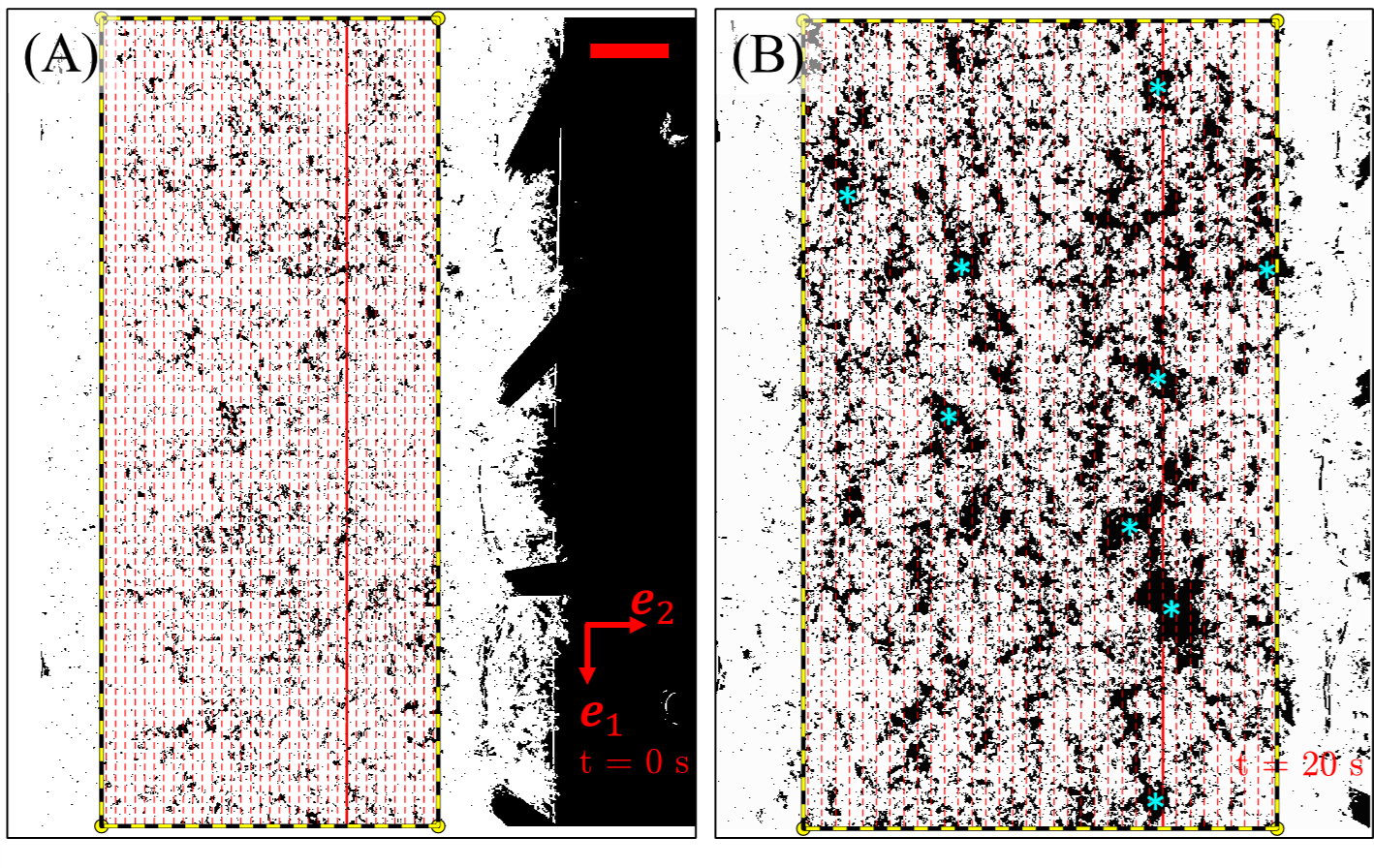}{Annotated image analysis of cyclic loading. }{\textbf{(A-B)} Binary image analysis of an ant raft, in which the ants are depicted white while gaps in the raft are depicted black, during cyclic loading at times \textbf{(A)} t=0 s and \textbf{(B)} t=20 s. Dotted red lines denote probed cross-sectional slices for which the cross-sectional length of ants is measured. The solid red line denotes the probed cross-section with the minimum value of $L_x$. The red scale bar in \textbf{(A)} represents $10 \ell$. Blue asterisks in \textbf{(B)} denote the centroids of voids. \label{SI: Cyclic image analysis}}{0.85}

\figuremacro{H}{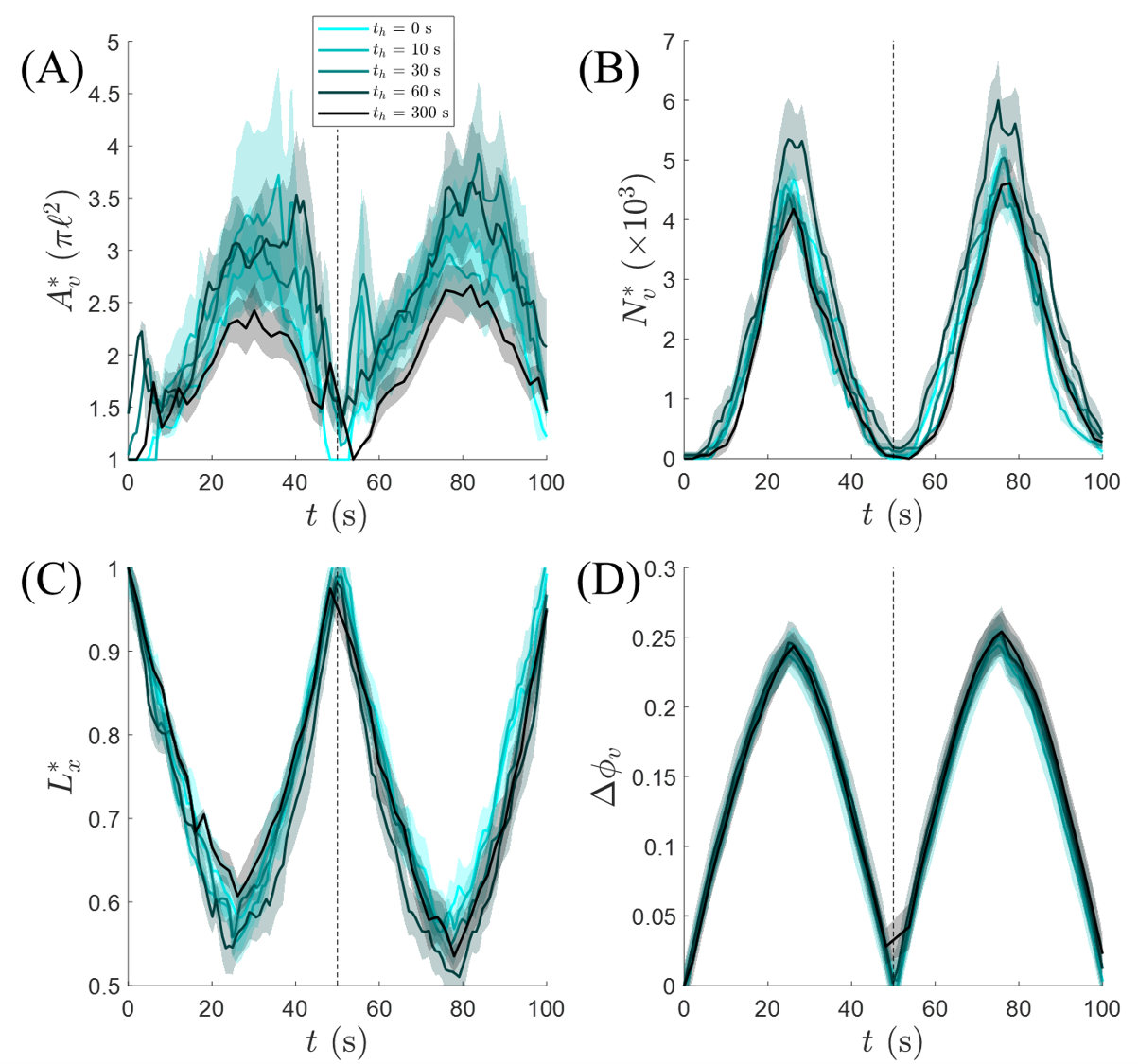}{}{\textbf{Cyclic damage - slower rate. } Ensemble-averaged \textbf{(A)} void area, $A_v^*$ (in units of $\pi \ell^2$), \textbf{(B)} number of voids, $N_v^*$ (in units of $10^{-3}$ voids per ant), \textbf{(C)} minimum cross sectional length, $L_x^*$, and \textbf{(D)} change in areal free volume, $\phi_v = \phi-\phi_0$ are provided for the cyclic loading data at 120 mm min$^{-1}$. ($W=1.2$). \label{SI: Cyclic damage response - 120}}{0.7}

\figuremacro{H}{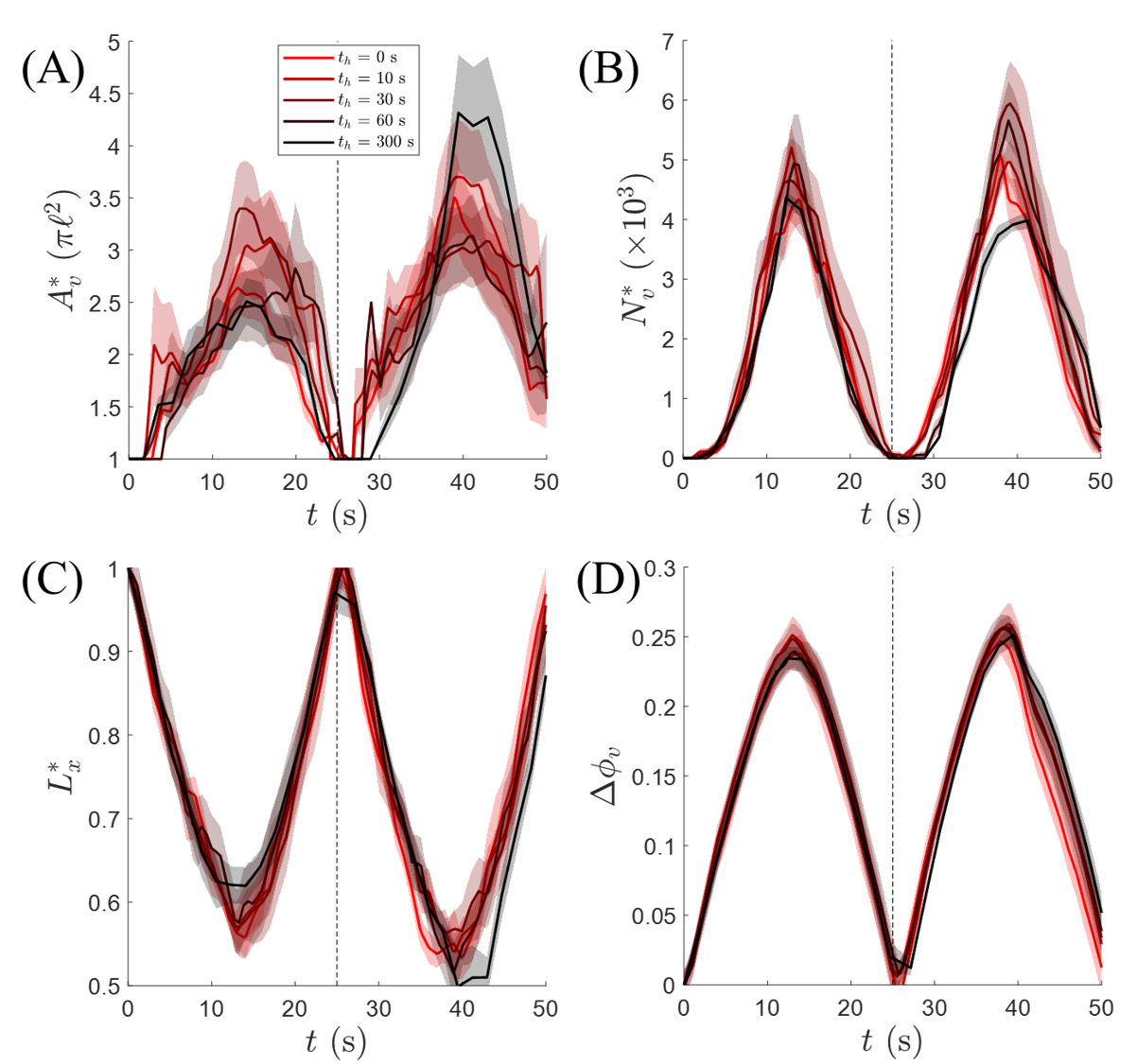}{}{\textbf{Cyclic damage - slower rate. } Ensemble-averaged \textbf{(A)} void area, $A_v^*$ (in units of $\pi \ell^2$), \textbf{(B)} number of voids, $N_v^*$ (in units of $10^{-3}$ voids per ant), \textbf{(C)} minimum cross sectional length, $L_x^*$, and \textbf{(D)} change in areal free volume, $\phi_v = \phi-\phi_0$ are provided for the cyclic loading data at 240 mm min$^{-1}$. ($W=2.4$). \label{SI: Cyclic damage response - 240}}{0.7}

\end{document}